\documentclass[nofootinbib,prl,twocolumn,showpacs,notitlepage,amsmath,amssymb,superscriptaddress,longbibliography]{revtex4-1} %

\usepackage[colorlinks=true,citecolor=blue,linkcolor=blue,urlcolor=blue]{hyperref}
\usepackage{cleveref}[2012/02/15]
\crefformat{footnote}{#2\footnotemark[#1]#3}

\newcommand{\sap}[1]{\textcolor{red}{[SP] #1}}

\usepackage{amsmath,amssymb,bm}

\usepackage{tikz}
\usetikzlibrary{external}
\usetikzlibrary{decorations.pathreplacing}
\usetikzlibrary{decorations.text}
\usetikzlibrary{decorations.pathmorphing}
\usetikzlibrary{shapes}

\definecolor{newblue}{RGB}{112,178,255}
\definecolor{neworange}{RGB}{255,204,112}
\definecolor{blue2}{RGB}{120,0,255}
\definecolor{red2}{RGB}{255,0,120}
\definecolor{green2}{RGB}{0,130,130}

\tikzset{snake it/.style={decorate, decoration={snake,segment length=1mm, amplitude=0.5mm}}}

\definecolor{darkred}{RGB}{245,186,183}
\definecolor{lightred}{RGB}{249,217,215}

\usetikzlibrary{arrows,calc,external,shapes.geometric}

\tikzset{
	>=stealth',
	help lines/.style={dashed, thick},
	important line/.style={thick},
	connection/.style={thick, dotted},
}

\tikzstyle{A}=[circle,draw=red!50,fill=red!20,thick]
\tikzstyle{R}=[circle,draw=blue!50,fill=blue!20,thick]
\tikzstyle{U}=[circle,draw=green!50,fill=green!20,thick]
\tikzstyle{V}=[circle,draw=orange!50,fill=orange!20,thick]

\tikzstyle{bag} = [align=center]

\newcommand{\eeq}{\end{equation}}
\newcommand{\ea}{\end{array}}

\def\eea{\end{eqnarray}}

\def\<{\langle}
\def\>{\rangle}

\def\bZ{\mathbb{Z}}

\def\cO{\mathcal{O}}

\def\bzz{\mathbb{Z}_3 \times \mathbb{Z}_3}

\usepackage{amsmath}
\usepackage{comment}
\usepackage{amssymb}
\usepackage{amsthm}

\theoremstyle{definition}

\makeatletter
\renewcommand\onecolumngrid{
\do@columngrid{one}{\@ne}%
\def\set@footnotewidth{\onecolumngrid}
\def\footnoterule{\kern-6pt\hrule width 1.5in\kern6pt}%
}
\makeatother

\begin{document}

\title{Boundary deconfined quantum criticality at transitions\\between symmetry-protected topological chains}

\author{Saranesh Prembabu}
\affiliation{Department of Physics, Harvard University, Cambridge, Massachusetts 02138, USA}

\author{Ryan Thorngren}
\affiliation{Kavli Institute of Theoretical Physics, University of California, Santa Barbara, California 93106, USA}

\author{Ruben Verresen}
\affiliation{Department of Physics, Harvard University, Cambridge, Massachusetts 02138, USA}

\begin{abstract}
Decades of research have revealed a deep understanding of topological quantum matter with protected edge modes.
We report that even richer physics emerges when tuning between two topological phases of matter whose respective edge modes are incompatible.
The frustration at the edge leads to novel boundary physics, such as symmetry-breaking phases with exotic non-Landau transitions---even when the edge is zero-dimensional.
As a minimal case study we consider spin chains with $\mathbb{Z}_3 \times \mathbb{Z}_3$ symmetry, exhibiting two nontrivial symmetry-protected topological (SPT) phases. At the bulk 1+1D critical transition between these SPT phases, we find two stable 0+1D boundary phases, each spontaneously breaking one of the $\bZ_3$ symmetries. Furthermore, we find that a single boundary parameter tunes a non-Landau boundary critical transition between these two phases. This constitutes a 0+1D version of an exotic phenomenon driven by charged vortex condensation known as deconfined quantum criticality. This work highlights the rich unexplored physics of criticality between nontrivial topological phases and provides insights into the burgeoning field of gapless topological phases.

\end{abstract}

\date{\today}

\maketitle

Symmetry-protected edge modes in phases of matter are well-understood when there is a finite energy gap to creating excitations in the bulk \cite{Gu09,chen_symmetry_2013,Senthil_2015}. For instance, in 1D systems \cite{pollmann_entanglement_2010,Turner11,Fidkowski_2011,chen_complete_2011,Schuch11} this leads to topologically protected ground state degeneracies which are exponentially localized near the endpoints \cite{affleck1988,Kennedy90}. 
However, edge modes at phase transitions and criticality \cite{Kestner11,Cheng11,Fidkowski11longrange,Sau11,Ruhman12,Grover12,Kraus13,Ortiz14,Keselman15,Ruhman15,Kainaris15,Iemini15,Lang15,Ortiz16,Montorsi17,Wang17,Ruhman17,Scaffidi17,Guther17,Kainaris17,Jiang18,Zhang18,Verresen18,Parker18,Keselman18,Chen18,Verresen21gapless,Duque21,Balabanov21,Chang22,Fraxanet22,Balabanov22} remain a fertile area of study.
Although such edge modes delocalize and disappear at phase transitions to the \emph{trivial} phase [Fig.~\ref{fig:bdyphasediagram}(a)] \cite{Tsui15,Tsui17,Bultinck19,Dupont21,Dupont21b, Tantivasadakarn21,pivots}, 
 it has been realized that transitions to \emph{other} phases---such as spontaneous symmetry-breaking phases [Fig.~\ref{fig:bdyphasediagram}(b)]\cite{Suzuki71,Briegel01,Son11, Scaffidi17,Parker18b,Verresen21gapless}---can leave part of the edge mode intact.

This raises a question which is fundamental for understanding the interplay between topology and criticality: What is the fate of edge modes when the system  transitions from one topological phase to \textit{another} nontrivial topological phase? To what extent do edge modes survive at the critical point?
  While previous work has studied this question in the non-interacting fermion case  \cite{Verresen18,verresen2020topology,Verresen21gapless}, here we explore a more generic and richer framework. Importantly, in our work, the edge modes of one phase are incompatible with those of the other phase due to differences in how they realize the symmetry action. The resulting frustration gives rise to fascinating boundary effects when both types of edge modes are forced to coexist and compete at criticality.

\begin{figure}[h!]
    \centering
    \begin{tikzpicture}
    \node at (2.5,3.3) {\includegraphics[scale=0.26]{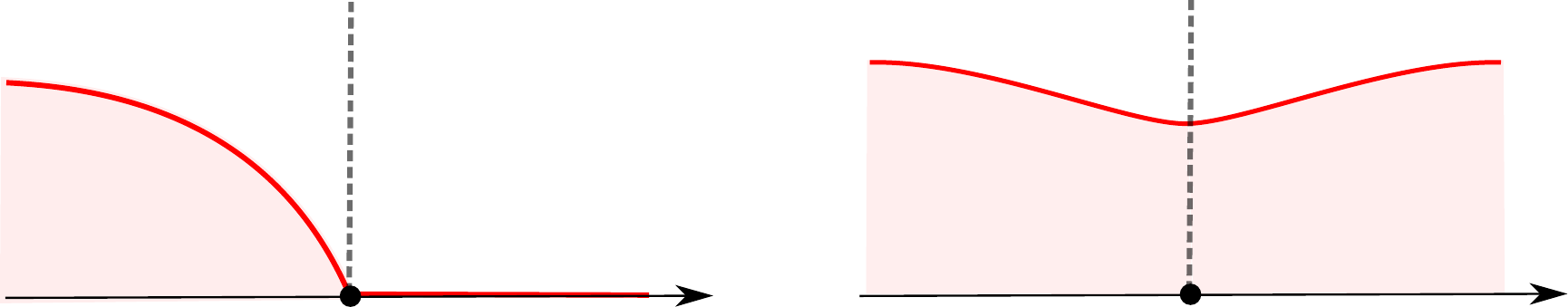}};
    \node at (-0.4,4) {SPT};
    \node at (0.3,2.3) {bulk tuning};
    \node at (1.2,4) {trivial};
    \node at (3.5,4) {SPT};
    \node at (4.4,2.3) {bulk tuning};
    \node at (5.2,4) {SSB};
    \node at (0,0) {\includegraphics[scale=0.26]{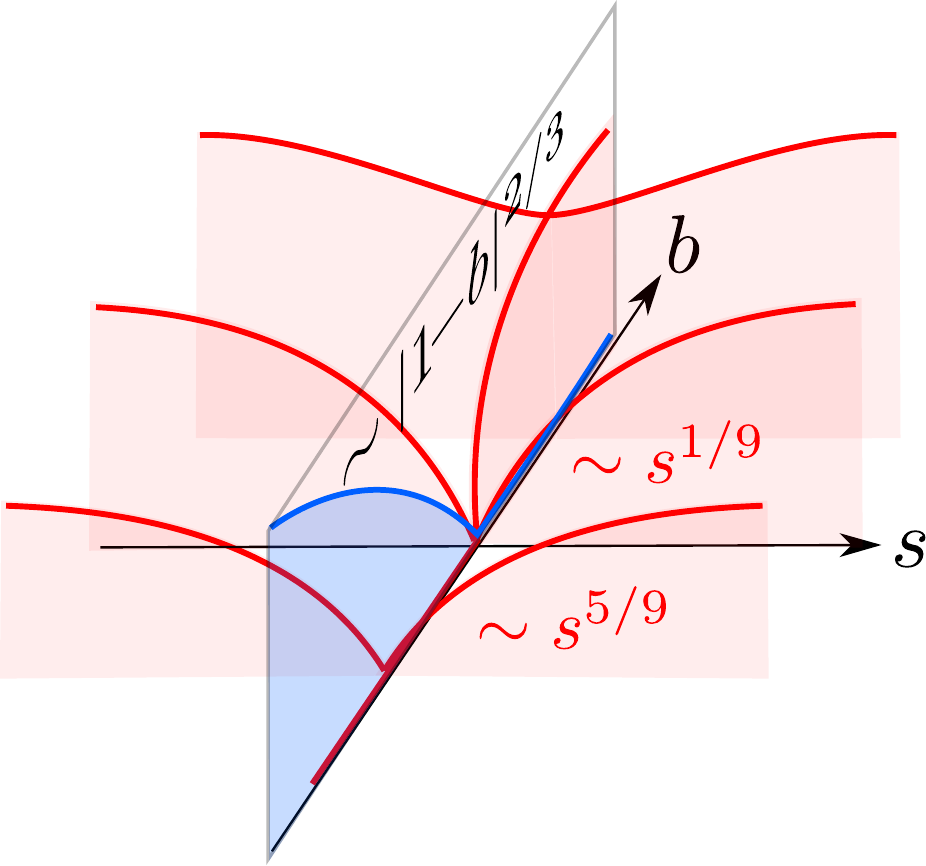}};
    \node at (4.4,0.2) {\includegraphics[scale=0.27,trim={0 2cm 0 0},clip]{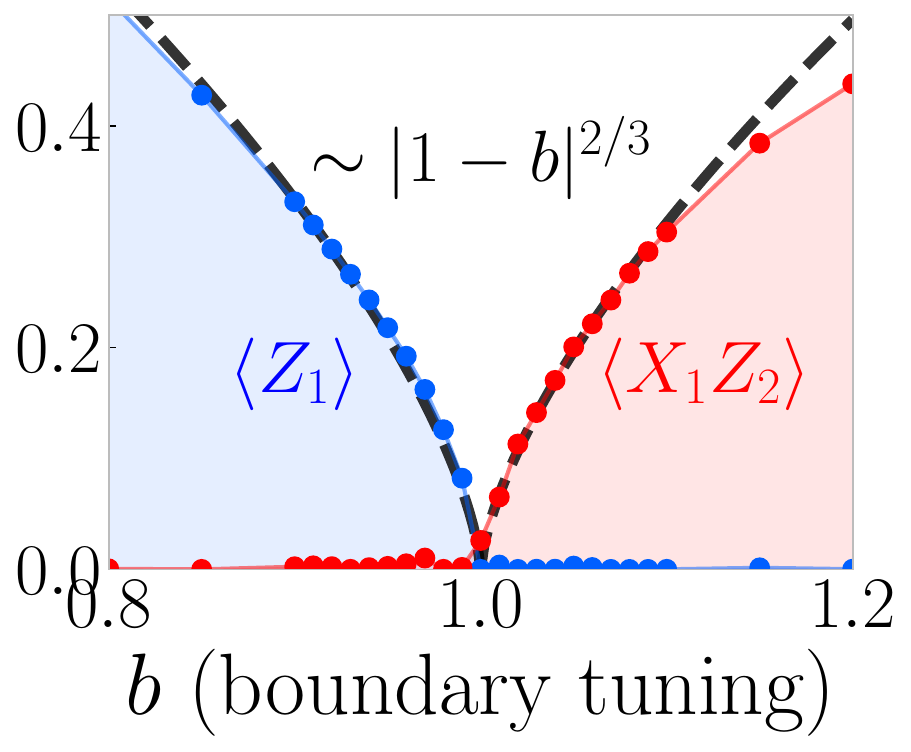}};
    \node at (4.5,-1.6) {$b$ (boundary tuning)};
    \node at (-1.8,4.1) {(a)};
    \node at (2.3,4.1) {(b)};
    \node at (-1.8,1.6) {(c)};
    \node at (2.3,1.6) {(d)};
    \draw[->,dashed,opacity=0.5] (0,-1.2) to[out=-20,in=190] (2.4,-1.4);
    \node at (-0.4,1.5) {SPT};
    \node at (1.4,1.5) {SPT'};
    \node at (-1.5, 3.3)[rotate = 90]{\tiny bdy. order param.};
    \end{tikzpicture}
    \caption{\textbf{SPT edge modes, criticality, and boundary DQCP.} Panels highlight what happens to SPT edge modes when tuning to quantum criticality. Based on end-to-end long-range order\cref{footnote:LRO} 
    of boundary order parameters: (a) tuning to trivial phase destroys edge modes and (b) edge modes can persist upon tuning toward a nontrivial phase, like a symmetry-breaking phase \cite{Scaffidi17,Parker18b,Verresen21gapless}. (c) In this work, we show a richer phenomenology at a transition between two distinct SPT phases protected by $\bZ_3 \times \bZ_3$ symmetry [Eq.~\eqref{eq:Hsb}]; there are distinct symmetry-breaking boundary conditions at criticality. (d) Moreover, there is a direct continuous transition (``DQCP'') between these two by tuning a boundary parameter.
    }
    \label{fig:bdyphasediagram}
\end{figure}
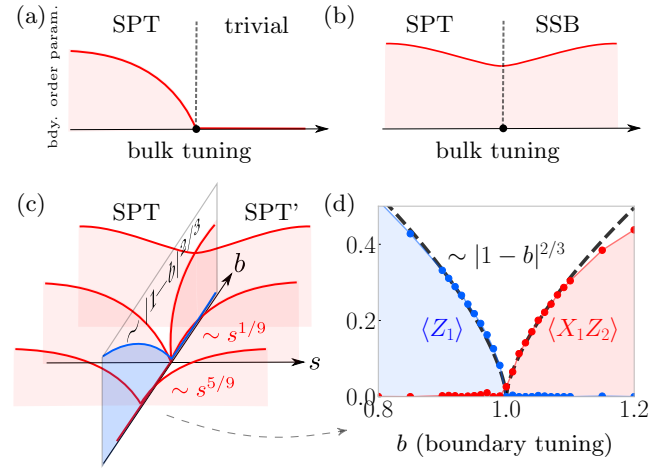

As a minimal example of this scenario, we study a transition between two $\bZ_3 \times \bZ_3$-symmetric spin chain Hamiltonians \cite{geraedtsmotrunich}, each phase hosting protected edge modes transforming under \emph{distinct} projective representations \cite{pollmann_entanglement_2010}. These two gapped phases are simple examples of the more general phenomenon of symmetry-protected topological (SPT) phases.
We find that
edge degeneracy typically persists at the critical point in two possible ways;
more precisely, there are two conformal boundary conditions each spontaneously breaking one 
of the two $\mathbb Z_3$ symmetries [Fig.~\ref{fig:bdyphasediagram}(c)]. These can be thought of as spontaneous symmetry-breaking phases in \textit{zero} spatial dimensions.
Moreover, we find a direct continuous boundary transition between these two [Fig.~\ref{fig:bdyphasediagram}(d)], where one symmetry breaks exactly when another is restored. This is a stark violation of the conventional Landau paradigm of phase transitions which posits that symmetry subgroups only break one at a time. In fact, it is a
0+1D
manifestation of a deconfined quantum critical point (DQCP), an exotic phenomenon
originally proposed for 2+1D \cite{Senthil04,Senthil04b,Levin_2004,Balents05, Vishwanath04, Ghaemi06, Sandvik07b, Grover07, Melko08, Sandvik10,KunChen13, Nahum15, Shao16, Wang_2017, Ma18, Ma19, Sreejith19, Li19, Takahashi20, Wang21, Ogino21} and 
recently explored in 1+1D \cite{Roberts19,Huang_2019,Mudry_2019,Jiang19, Sun19, Yang20, RJM21,ZhangLevin2022}. Indeed, we 
discuss how even the mechanism is quite similar to that in higher dimensions, namely, condensing defects for one symmetry-breaking order gives rise to long-range order for the other ~\cite{Levin_2004}.

\footnotetext{For gapped SPT phases its value depends on the choice of ground state. In contrast, gapless phases can
have robust 0+1D symmetry-breaking on the edge \cite{Scaffidi17,Parker18b,Verresen21gapless}.\label{footnote:LRO}}

Moreover, we show that the bulk critical point itself has a nontrivial topological invariant---making it an instance of gapless SPT or symmetry-enriched criticality \cite{Verresen21gapless}.
The conventional lore for topologically nontrivial SPT phases, shown rigorously in the gapped case, is that edge modes are guaranteed by a bulk-boundary correspondence. However, at the boundary critical point, we report here, edge modes disappear. This shows that the notion of bulk-boundary correspondence is more subtle for gapless SPT phases, opening up exciting
future research directions.

\textbf{$\bZ_3 \times \bZ_3$ cluster SPT chains.---}
Define 
shift and clock matrices:
\begin{equation}
X = \begin{pmatrix} 0 & 0 & 1 \\ 1 & 0 & 0 \\ 0 & 1 & 0 \end{pmatrix} \quad \textrm{and} \quad Z = \begin{pmatrix}  1 & 0 & 0 \\ 0 & \omega & 0\\ 0 & 0 & \omega^2 \end{pmatrix},
\end{equation}
where $\omega= e^{2\pi i/3}$. We 
consider quantum chains respecting the $\bZ_3 \times \bZ_3$ symmetry generated on 
even and odd sublattices:
\begin{equation}
U^e = \prod_j X_{2j} \quad \textrm{and} \quad U^o = \prod_j X_{2j+1}.
\end{equation}
Following Ref.~\onlinecite{geraedtsmotrunich}, we define ``cluster Hamiltonians'' \cite{Briegel01} for two distinct non-trivial $\bZ_3 \times \bZ_3$ SPT phases 
\begin{equation}\label{eqnclusterhams}
\begin{gathered}
H_\omega = - \sum_j \big(Z_{2j-1} X_{2j} Z_{2j+1}^\dagger + Z_{2j} X_{2j+1}^\dagger Z_{2j+2}^\dagger + h.c. \big) \\
H_{\bar \omega} = - \sum_j \big( Z_{2j-1} X_{2j}^\dagger Z_{2j+1}^\dagger + Z_{2j} X_{2j+1} Z_{2j+2}^\dagger + h.c. \big).
\end{gathered}
\end{equation}
The effective low-energy action of $U^e$ and $U^0$ on
each boundary is such that they commute only up to a projective phase $\omega$ or $\bar \omega$, leading to a threefold degenerate ground state space (per edge) \cite{geraedtsmotrunich}. 

The edge projective symmetry action is detectable via bulk string order parameters.
That is, 
among operators of the form $\cdots X_{2j-6} X_{2j-4} X_{2j-2} \cO_{2j}$ 
(a $\bZ_3^e$-string operator), 
only those with $\bZ_3^o$-charged $\cO_{2j}$ have have long-range order (LRO) \cite{Pollmann12b}, and vice versa for the other symmetry. For instance, $H_\omega$ has LRO in an $\omega$-charged $\bZ_3^e$-string operator
\begin{equation}
\lim_{|k-j| \to \infty} \langle Z_{2j-1} X_{2j} X_{2j+2} \cdots X_{2k} Z_{2k+1}^\dagger \rangle = 1, \label{eq:sop}
\end{equation}
while $H_{\bar \omega}$ has an $\bar \omega$-charged $\bZ_3^e$-string operator. While the left hand side of Eq.~\eqref{eq:sop} is unity only for the fixed-point Hamiltonian $H_\omega$, it remains nonzero throughout the SPT phase \cite{Pollmann12b}.

\textbf{Numerical method.---} We 
confirm our 
CFT analysis 
using 
density matrix renormalization group (DMRG) simulations \cite{White92,Hauschild18} on finite chains of lengths $25 \lesssim L \lesssim 125$. At each length, we considered the limit of bond dimesnion $\chi \to \infty$, with simulations run up to $\chi = 170$ found to 
sufficiently guarantee convergence for 
ground state end-to-end correlators and excited state energy levels. 
For technical efficiency, instead of implementing the cluster Hamiltonian directly, we simulated a unitarily-equivalent three-state Potts chain as described in Ref.~\cite{suppl}.

\textbf{Criticality and boundary symmetry breaking.---}We study a linear interpolation between the two non-trivial cluster Hamiltonians \eqref{eqnclusterhams}:
\begin{equation}
H(s,b) = (1+s)H_\omega + (1-s) H_{\bar \omega} - b(X_1 + X_{2N+1} + h.c.). \label{eq:Hsb}
\end{equation}
Since we are interested in 
edge behavior, we have 
open boundary conditions 
with $j \in [1,2N+1]$
and 
boundary tuning parameter $b$ to explore generic boundary behavior. 

This model exhibits a direct transition at the midpoint 
$s = 0$. In fact, 
a local unitary (the SPT entangler \cite{Santos15}) maps  $H_1 \mapsto H_\omega \mapsto H_{\bar \omega} \mapsto H_1$, where $H_1 = - \sum_j X_j + \textrm{H.c.}$ is a trivial phase. So the bulk critical point can be mapped to one between the trivial and SPT phase, $H_1 + H_\omega$, which 
in Ref.~\onlinecite{Tsui17}
was found to be be described by a certain orbifold of two copies of the three-state Potts conformal field theory (Potts$^2$ CFT). However, these entangler transformations do not apply for open boundary conditions, and we will find 
$H_\omega + H_{\bar \omega}$ 
has much richer boundary criticality than $H_1 + H_\omega$; we will also discuss how 
a bulk symmetry-protected topological invariant detects this difference. We note that this $s=0$ critical point 
belongs to a one-parameter family of theories stabilized by $\bzz$, translation, and charge conjugation (see Refs. \cite{suppl,RJM21, Roberts_conversation, Whitsitt_2018} therein for details on adjacent bulk phases).

Unlike in gapped SPT phases, 
string operators 
[Eq.~\eqref{eq:sop}] no longer have LRO at 
criticality. Instead they decay algebraically with universal exponents 
distinguishing 
$H_\omega + H_{\bar \omega}$ and $H_1 + H_\omega$. For example, 
considering 
charges of the `lightest' string operators, i.e., those with the smallest such exponents, 
$H_\omega + H_{\bar \omega}$ has two degenerate $U^e$-string operators with $U^o$ charges $\{\omega,\bar \omega\}$ [e.g., the lattice string operator 
Eq.~\eqref{eq:sop}], while $H_1 + H_\omega$ has 
charges $\{1,\omega\}$; these correspond to string operators 
with LRO in the nearby symmetric phases. This bulk topological invariant proves that these two CFTs cannot be connected by a $\bZ_3 \times \bZ_3$-symmetric path without passing through a multi-critical point or tuning off criticality.

In gapped SPTs, 
LRO of charged strings \eqref{eq:sop}
directly imply edge modes. 
Analogously, we might expect a similar bulk-boundary correspondence 
can distinguish the ``trivial" transition $H_1 + H_\omega$ from the ``topological" one $H_\omega + H_{\bar \omega}$. To explore this, we turn to a more concrete analysis of Eq.~\eqref{eq:Hsb}, using analytic and numerical methods inspired by 
Ref.~\onlinecite{Verresen21gapless}. 

In the fine-tuned case $b = 0$, 
zero mode operators $Z_1$ and $Z_{2N+1}$ 
commute with $H$ [Eq.~\eqref{eq:Hsb}]. Their $\bZ_3^o$ charge implies a threefold 
degenerate spectrum. Morally, $Z_1$ and $Z_{2N+1}$ are order parameters for a spontaneous symmetry-breaking (SSB) boundary
and are indeed LRO in time. They are also phase-locked 
across the critical bulk: $\langle Z_1 \rangle = \langle Z_{2N+1}\rangle$ in all three ground states (i.e., unlike 
in gapped SPTs, 
degeneracies are not independent for both edges). We call this boundary phase ``$o$-SSB.'' Note that 
bulk gaplessness is what ensures a well-defined 
boundary SSB; in contrast, in 
gapped SPT phases 
end-to-end LRO requires a certain 
basis of degenerate ground states. Indeed, 
gapped SPT edge modes, 
being genuine zero-dimensional systems, 
have no robust notion of ``phase of matter.''

Adding nonzero $b$ 
splits 
degeneracy for finite systems, similar to the exponentially small finite-size splitting of gapped SPTs 
\cite{Kennedy90,kitaev_unpaired_2001}.
At criticality ($s=0$), 
edge modes split \emph{algebraically} $\sim 1/N^\alpha$ with 
boundary-condition-dependent exponent $\alpha$. 
Crucially $\alpha>1$, such that 
degeneracy is relative to 
\emph{bulk} finite-size splitting $\sim 1/N$ ~\cite{Verresen21gapless}. We numerically confirm this faster-than-$1/N$ splitting, and hence 
boundary stability, in Fig.~\ref{fig:energies}(a). 
Later we derive a mapping to the Potts model, implying the universal exponent $\alpha = 5/3$.  

\begin{figure}
    \centering
    \begin{tikzpicture}
    \node at (0,0) {
    \includegraphics[scale=0.32]{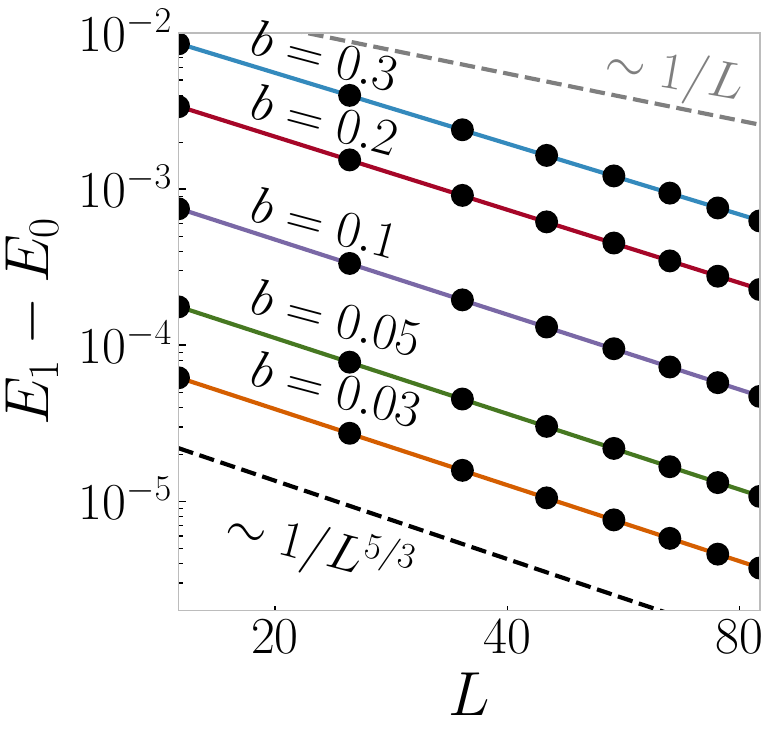}
    };
    \node at (4.5,0.25) {
    \includegraphics[scale=0.25]{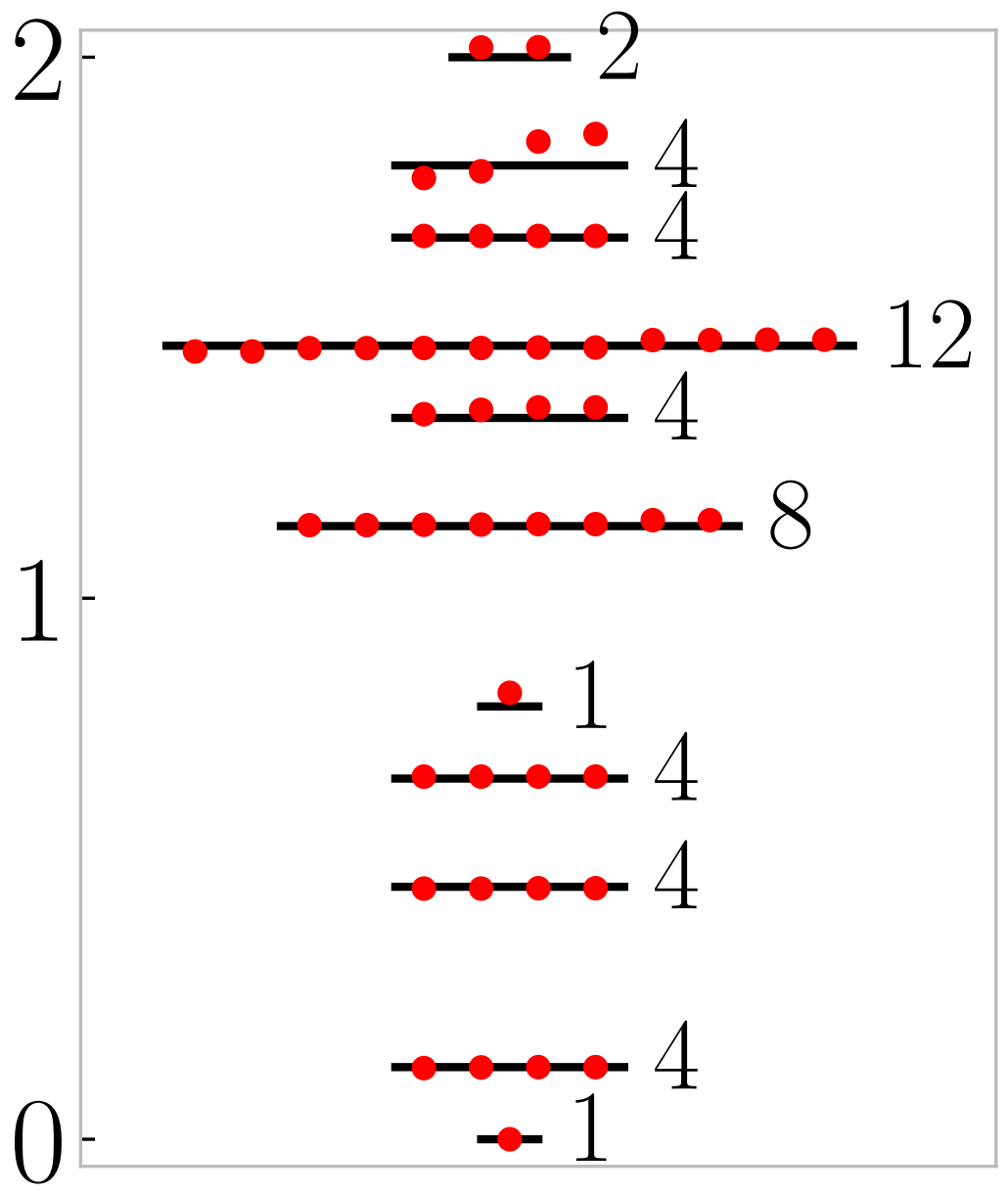}
    };
    \node at (2.6, 0.2)[rotate = 90]{ $(E_n-E_0)L/v$};
    \node at (-2,1.8) {(a)};
    \node at (2.5,1.8) {(b)};
    \end{tikzpicture}
    \caption{\textbf{Edge modes and boundary dissolution at SPT criticality.} We consider Eq.~\eqref{eq:Hsb} with open boundaries. (a) For $b<1$, the boundary spontaneously breaks $U^o$. This degeneracy's finite-size splitting matches the 
    CFT prediction $\sim L^{-5/3}$. 
    Edge modes become exactly degenerate in the CFT limit. (b) At $b=1$, the boundary undergoes a transition between two distinct symmetry-breaking phases. Here, we find a unique ground state.  Red dots denote the numerically extracted universal finite-size spectrum (for $L=25$; here $v = 3\sqrt{3}\pi$). Remarkably, this 
    matches the spectrum of 
    Potts CFT \emph{without boundaries} (black lines). This signifies that at this point, the 
    bulk-boundary distinction is blurred (see main text).}
    \label{fig:energies}
\end{figure}

The other easy limit, $b \to \infty$, 
projects $X_1 = X_{2N+1}=1$, 
i.e. 
throws out 
sites $1$ and $2N+1$ and 
operators acting on them. 
One has the same model as when $b=0$, but with $j \in [2,2N]$. The story 
above repeats, except now $U^e$ is spontaneously broken at the boundary, not $U^o$. This is a distinct boundary phase (``$e$-SSB'') from $b = 0$, 
raising the question of what 
boundary transition 
occurs as we tune the boundary coupling.

We note 
this perturbatively stable boundary symmetry breaking
requires the exotic topological nature of the model $H_\omega + H_{\bar{\omega}}$.  For comparison, the trivial gapless theory $H_1 + H_\omega$ has a generically nondegenerate conformal spectrum and no boundary symmetry breaking except at some unstable fine-tuned boundary points. The difference lies in 
the so-called \emph{boundary disorder operators}, i.e., operators 
toggling between 
superselection sectors of 
0+1D SSB ground states.
For $H_1 + H_\omega$, 
perturbing 
a fine-tuned degenerate edge with infinitesimal $X_1$ will disorder the 0+1D SSB and flow to a unique symmetry-preserving edge. In contrast, $H_\omega+ H_{\bar \omega}$ has no RG-relevant symmetry-allowed boundary disorder operator with which we can perturb the edge. This intuitively matches the
bulk topological invariants 
and also follows from 
CFT (explained in Refs~\cite{suppl,cardyBCFT}).
The relevant perturbation, shown in Table~\ref{tab:order_disorder_operators}, carries nontrivial charge under the unbroken symmetry and thus cannot be generated under RG!

Finally we remark that the boundary order parameters and disorder operators for the gapless regime match localized projective symmetry generators from the adjacent gapped SPT phases. Although the gapped SPT phase is agnostic with respect to automorphisms of  $\mathbb{Z}_3\times \mathbb{Z}_3$,
 the gapless theory selects a specific choice of projective symmetry generators to play the role of boundary order parameter or disorder operator. This physically corresponds to the fact that the boundary of the gapless phase has genuine 0+1d SSB, in contrast to the edge of a gapped SPT phase.

\textbf{DQCP in zero dimensions.---}To recap, for $b \approx 0$, $H_\omega + H_{\bar \omega}$ with open boundaries spontaneously breaks the odd-sublattice $\bZ_3$ symmetry, 
while for $b \to \infty$ it breaks the even one. 
It turns out 
these two phases persist for all $b$, \emph{except at $b = 1$}, where there is a direct transition. This boundary transition is continuous, and both symmetries are unbroken there. Indeed, for $b = 1$, we find \emph{no} ground state degeneracy (Fig.~\ref{fig:energies}), contrary to a naive expectation from the bulk topological invariant.

Tuning left and right boundary couplings 
simultaneously (see Eq.~\eqref{eq:Hsb}) lets us use order parameters' end-to-end correlations to detect the transition, which occurs independently on both edges. In particular $\langle Z_1 Z^\dagger_{2N+1} \rangle$ is nonzero in the $o$-SSB boundary phase $(0 \le b < 1)$ and zero in the $e$-SSB boundary phase $(b >1)$, and vice versa for $ \langle X_1 Z_2 Z^\dagger_{2N} X^\dagger_{2N+1}\rangle$. The square root gives the boundary vacuum expectation value (vev). Using DMRG, 
we obtain Fig.~\ref{fig:bdyphasediagram}(d) and clearly see the direct continuous transition at $b = 1$. 
Later, we analytically show both vevs vanish at $b = 1$ 
with unbroken symmetry and no ground state degeneracy.

This continuous SSB-to-SSB transition resembles 
deconfined quantum criticality points (DQCP) in higher dimensions. A key feature of DQCP is that the ``vortex" in one ordered phase is charged under the symmetry broken in the other. Thus they cannot simultaneously condense, leading to a Landau-forbidden transition. The same mechanism prevails here, with the role of 
vortices played by 
relevant boundary disorder operators of Table ~\ref{tab:order_disorder_operators}.
Another salient DQCP feature is an emergent symmetry exchanging nearby SSB phases, which we will show indeed occurs at $b=1$.
Despite these similarities,
an anomalous symmetry is arguably missing. Indeed, a \emph{bona fide} zero-dimensional anomaly is usually understood to be a projective representation; since this implies 
degeneracy, it cannot be present at $b=1$. 
Thus, following Ref.~\onlinecite{RJM21}, we use the term DQCP in a slightly broader context, namely, a non-Landau transition between distinct SSB phases stabilized by 
condensing 
charged defect operators.

\footnotetext{These are lattice expressions for the boundary disorder operators at the extreme limits $b=0$ and $b=\infty$; moreover, for generic $b$ their expansion in continuum field is dominated by a boundary disorder operator. We also mention that $Z_1$ is identically zero at $b=\infty$, but for finite $b>1$ is in the same universality as $X_2 Z_3$. \label{footnote:disorder}}

\begin{table}[]
    \centering
    \begin{tabular}{c||c|c}
         & Order operator &  Disorder operator\\ \hline \hline
    $o$-SSB  & \color{blue}$Z_1$& \color{blue} $X_1Z_2^{(\dagger)}$\\
    $e$-SSB & \color{red}$X_1Z_2^{(\dagger)}$ &\textcolor{red}{$Z_1$} (or \textcolor{red}{$X_2 Z_3^{(\dagger)}$})\\
    \end{tabular}
    \caption{The most relevant disorder operators of the odd symmetry-breaking boundary 
    ($b<1$) are 
    order parameters of the even symmetry-breaking boundary 
    ($b>1$) and vice versa. Here we show  
    left boundary lattice expressions\cref{footnote:disorder}.
    Restoring one symmetry requires condensing said disorder operator, thereby spontaneously breaking the other symmetry; this is the mechanism leading to the 0+1D `non-Landau' DQCP.}
    \label{tab:order_disorder_operators}
\end{table}

We numerically verify symmetry restoration and nondegeneracy at $b=1$ by computing a finite chain's spectrum, 
Fig.~\ref{fig:energies}(b). 
Remarkably, this spectrum coincides with the known analytic result for a \emph{single} Potts chain with \emph{periodic}%
(twisted)
boundary conditions \cite{PetkovaZuber}. 
This is no coincidence, as we 
now demonstrate.

\textbf{Mapping to single Potts chain.---}Remarkably, the open chain in Eq.~\eqref{eq:Hsb} is unitarily equivalent to a single three-state Potts chain on a ring with some defects depending on $b$. 
The mapping is summarized in Fig.~\ref{fig:mapping} (further details are provided in 
the Supplemental Material \cite{suppl}).
The $\bZ^e_3$ physical symmetry is the global $\bZ_3$ symmetry of the Potts ring, while the eigenvalues of the $\bZ^o_3$ generator label the $\bZ_3$ twisted boundary conditions of the Potts ring. The result is that $b$ tunes the strength of a single exchange term on opposite sides of the ring. 
The DQCP at $b=1$ corresponds to translation symmetry,  
where 
the spectrum matches that of the Potts chain on a ring, which is nondegenerate.

With this mapping, 
dominant boundary operators 
are identified through the Potts 
defect conformal field theory \cite{Ginsparg88,francesco_conformal_1997, Kormos_2009, potts_bcft98, potts_dcft22, Fendley_2009}
confirming our 
claims in Fig. ~\ref{fig:bdyphasediagram}. 
For example, at $b<1$, the dominant symmetry-allowed boundary perturbation corresponds to an irrelevant $\psi^A\psi^{B\dagger}$ CFT operator of dimension $4/3$ coupling the two 
chains' endpoints (Fig. \ref{fig:mapping}), while the $\bZ^e_3$-charged disorder operator corresponds to $\psi^A$ of dimension $2/3$ leading to the $|s|^{5/9}$ scaling of Fig. \ref{fig:bdyphasediagram}(c).
Similary, at the DQCP, scalings of Figs. \ref{fig:bdyphasediagram}(c) and \ref{fig:bdyphasediagram}(d) arise from the $2/15$ dimensional 
$\bZ^e_3$ ($\bZ^o_3$) charged boundary operators $\sigma \bar{\sigma}$ ($\mu \bar{\mu}$) and the $4/5$ dimensional 
the symmetric boundary perturbation $\epsilon \bar{\epsilon}$.

\begin{figure}
    \centering
     \centering
    \begin{tikzpicture}
    \node at (1,5) {\includegraphics[width=6cm]{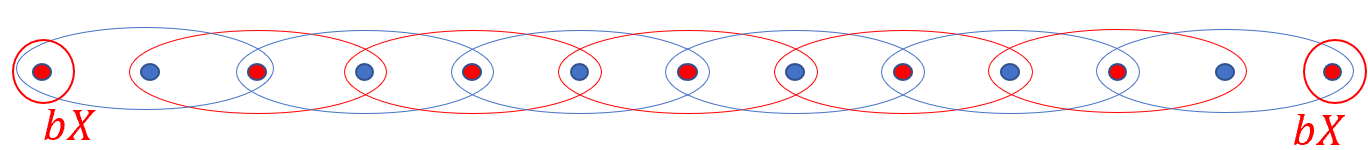}};
    \node at (-3.0,5) {Cluster};
     \draw[->,solid,opacity=1.0] (0.8,4.5) to (0.8,3.7);
     \node at (1-2.0, 4.1) {\textit{KW on odd sites}};
    \node at (1,3) {\includegraphics[width=6cm]{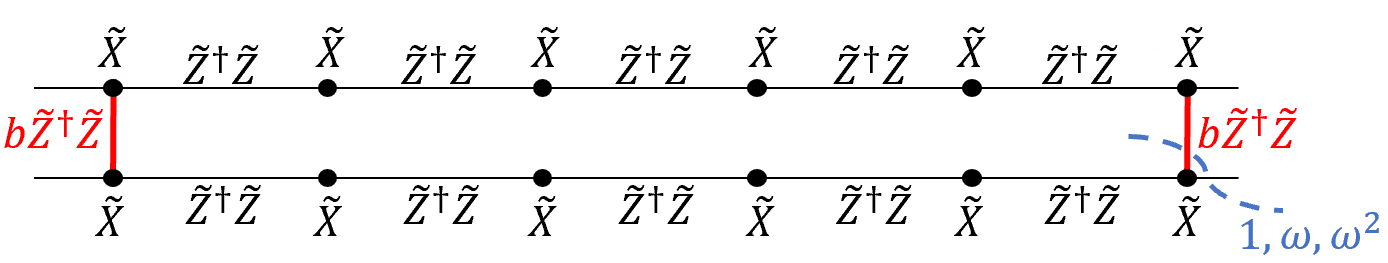}};
    \node[bag] at (-3.0,3) {Potts \\ $\times$ \\ Potts};
    \node at (1,0) {\includegraphics[width=5cm]{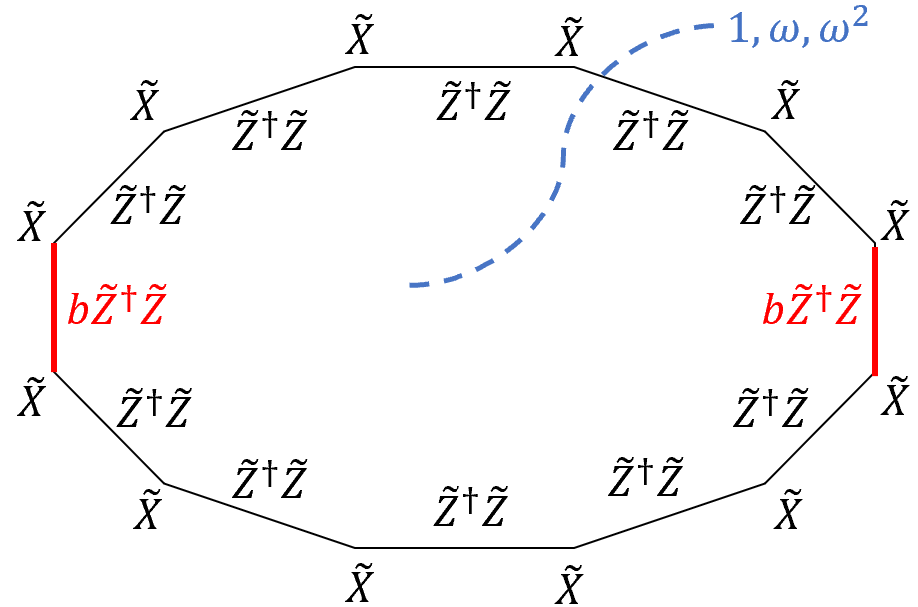}};
    \node at (1-1.0, 2.1) {\textit{Unfold}};
     \draw[->,solid,opacity=1.0] (1,2.5) to (1,1.7);
    \node[bag] at (-3,0,0) {Single \\ Potts};
     \node at (1+2.0, 4.1) { \scriptsize $\tilde{Z}_{2j}^A = Z_{2j} X_{2j-1} X_{2j-3} X_{2j-5} \ldots $};

    \end{tikzpicture}
    \caption{\textbf{Mapping critical $\bZ_3\times \bZ_3$ cluster chain with boundaries to a single Potts chain.} There is an exact unitary map from a finite open cluster chain to a finite \textit{closed} Potts chain with defects. First we apply a Kramers Wannier transformation on odd sites and appropriately parametrize the resulting even sites to have the form of two $c=4/5$ Potts chains only coupled at their boundaries by the boundary perturbation. Then we ``unfold'' this doubled system by simply viewing it as a single Potts system on a closed loop with defects and twisted sectors.
    }
    \label{fig:mapping}
\end{figure}
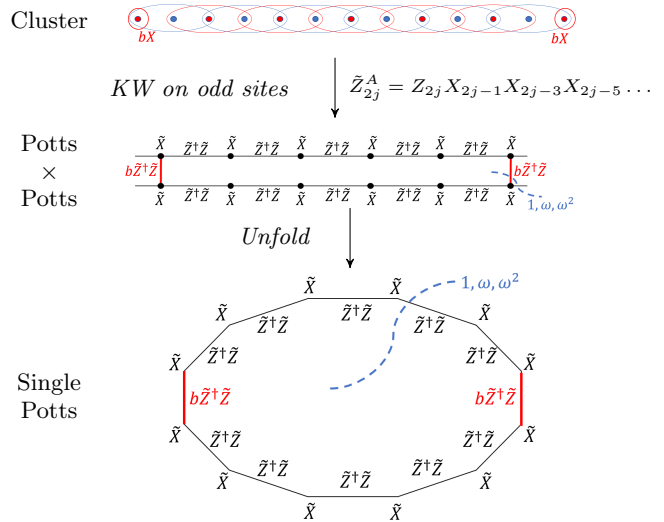

Furthermore, the Potts chain's Kramers-Wannier duality interchanges  $U^e$ and $U^o$ symmetries and all order and disorder operators. It  sends $b = 1 + \delta b$ to $b = 1-\delta b$ for $\delta b \ll 1$, acting as an emergent duality in the boundary phase diagram.
Two such transformations 
translate the Potts ring. At $b=1$, the  emergent translation symmetry relates boundary degrees of freedom to bulk degrees of freedom. Thus the boundary critical point is also a ``delocalized" QCP.

\textbf{Outlook.---}
We studied a minimal example of competition between two inequivalent types of topologically-protected edge mode with $\mathbb{Z}_3 \times \mathbb{Z}_3$ symmetry. We found that as a result of this competition, there are effectively fewer edge modes at criticality, and they organize themselves into one of two distinct boundary-symmetry-breaking phases breaking only a three-fold subgroup. Most strikingly, there is an unconventional direct boundary transition between these two symmetry-breaking regimes. At this boundary transition, edge modes disappear and emergent features of a deconfined quantum critical point appear. These results were obtained using conformal field theory and tensor network simulations on a critical one-dimensional open-chain lattice model on an open chain with a $\mathbb{Z}_3 \times \mathbb{Z}_3$ symmetry.

SPT transitions and edge modes of gapless systems merit further study. 
Our results encourage exploring other direct transitions between nontrivial SPT phases, where, as we have exemplified, novel boundary physics is expected.
Examples include 
boundaries of the $\mathbb Z_2 \times \mathbb Z_2 \times \mathbb Z_2^T$ SPT-SPT' transition in Eq.(28) of Ref.~\onlinecite{Verresen21gapless}
and the $c=2$ multicritical point where all three  $\bZ_3 \times \bZ_3$ SPT phases meet.
Another major open question regards bulk-boundary correspondence
for gapless SPT phases. 
Remarkably we have found that even with 
a nontrivial bulk topological invariant, boundary edge modes can disappear in a boundary DQCP. It remains unknown how general this phenomenon is.
Insights might also be gained by understanding 
boundary conditions
as RG flows to 1+1D gapped phases \cite{Cho17b,Cho17c,Cardy17}.
Another open question is to explore higher-dimensional analogs,
such as transitions between nontrivial
2+1D $\bZ_n$ SPTs.

Finally, it would be exciting to explore these phenomena in experiment. Intriguingly, $\bZ_3 \times \bZ_3$ SPT phases have been predicted in optical lattices of cold alkaline-earth atoms \cite{Fromholz_2019, Capponi_2020}. While numerical simulations found a direct first-order transition between the two nontrivial SPT phases, our work suggests a broader phase diagram can have a direct continuous transition, where one would observe 0+1D boundary DQCP. To facilitate such experimental explorations, one can map the three-body cluster Hamiltonian to a two-body interacting system \cite{future_qt}, similar to what has been done for the $\mathbb{Z}_2$ case \cite{Verresen17}.

\textbf{Acknowledgments.---}The authors thank Hart Goldman, Zohar Komargodski, Patrick Ledwith, Max Metlitski, Brenden Roberts, Rahul Sahay, Rhine Samajdar, Yifan Wang, Carolyn Zhang  and Ashvin Vishwanath for stimulating conversations, and the latter also for detailed comments on the manuscript. SP also thanks Jayalakshmi Namasivayan for support.
DMRG simulations were performed on the Harvard FASRC facility using the TeNPy Library \cite{Hauschild18}, which was inspired by a previous library \cite{Kjaell13}. SP was supported by the National Science Foundation
Graduate Research Fellowship under Grant No. 1745303. RV is supported by the Harvard Quantum Initiative Postdoctoral Fellowship in Science and Engineering, and by the Simons Collaboration on Ultra-Quantum Matter, which is a grant from the Simons Foundation (651440, Ashvin Vishwanath). RT is supported in part by the National Science Foundation under Grant No. NSF PHY-1748958.

\bibliography{PRL_bib.bib}

\begin{thebibliography}{108}%
\makeatletter
\providecommand \@ifxundefined [1]{%
 \@ifx{#1\undefined}
}%
\providecommand \@ifnum [1]{%
 \ifnum #1\expandafter \@firstoftwo
 \else \expandafter \@secondoftwo
 \fi
}%
\providecommand \@ifx [1]{%
 \ifx #1\expandafter \@firstoftwo
 \else \expandafter \@secondoftwo
 \fi
}%
\providecommand \natexlab [1]{#1}%
\providecommand \enquote  [1]{``#1''}%
\providecommand \bibnamefont  [1]{#1}%
\providecommand \bibfnamefont [1]{#1}%
\providecommand \citenamefont [1]{#1}%
\providecommand \href@noop [0]{\@secondoftwo}%
\providecommand \href [0]{\begingroup \@sanitize@url \@href}%
\providecommand \@href[1]{\@@startlink{#1}\@@href}%
\providecommand \@@href[1]{\endgroup#1\@@endlink}%
\providecommand \@sanitize@url [0]{\catcode `\\12\catcode `\$12\catcode
  `\&12\catcode `\#12\catcode `\^12\catcode `\_12\catcode `\%12\relax}%
\providecommand \@@startlink[1]{}%
\providecommand \@@endlink[0]{}%
\providecommand \url  [0]{\begingroup\@sanitize@url \@url }%
\providecommand \@url [1]{\endgroup\@href {#1}{\urlprefix }}%
\providecommand \urlprefix  [0]{URL }%
\providecommand \Eprint [0]{\href }%
\providecommand \doibase [0]{http://dx.doi.org/}%
\providecommand \selectlanguage [0]{\@gobble}%
\providecommand \bibinfo  [0]{\@secondoftwo}%
\providecommand \bibfield  [0]{\@secondoftwo}%
\providecommand \translation [1]{[#1]}%
\providecommand \BibitemOpen [0]{}%
\providecommand \bibitemStop [0]{}%
\providecommand \bibitemNoStop [0]{.\EOS\space}%
\providecommand \EOS [0]{\spacefactor3000\relax}%
\providecommand \BibitemShut  [1]{\csname bibitem#1\endcsname}%
\let\auto@bib@innerbib\@empty
\bibitem [{\citenamefont {Gu}\ and\ \citenamefont {Wen}(2009)}]{Gu09}%
  \BibitemOpen
  \bibfield  {author} {\bibinfo {author} {\bibfnamefont {Zheng-Cheng}\
  \bibnamefont {Gu}}\ and\ \bibinfo {author} {\bibfnamefont {Xiao-Gang}\
  \bibnamefont {Wen}},\ }\bibfield  {title} {\enquote {\bibinfo {title}
  {Tensor-entanglement-filtering renormalization approach and
  symmetry-protected topological order},}\ }\href {\doibase
  10.1103/PhysRevB.80.155131} {\bibfield  {journal} {\bibinfo  {journal} {Phys.
  Rev. B}\ }\textbf {\bibinfo {volume} {80}},\ \bibinfo {pages} {155131}
  (\bibinfo {year} {2009})}\BibitemShut {NoStop}%
\bibitem [{\citenamefont {Chen}\ \emph
  {et~al.}(2013{\natexlab{a}})\citenamefont {Chen}, \citenamefont {Gu},
  \citenamefont {Liu},\ and\ \citenamefont {Wen}}]{chen_symmetry_2013}%
  \BibitemOpen
  \bibfield  {author} {\bibinfo {author} {\bibfnamefont {Xie}\ \bibnamefont
  {Chen}}, \bibinfo {author} {\bibfnamefont {Zheng-Cheng}\ \bibnamefont {Gu}},
  \bibinfo {author} {\bibfnamefont {Zheng-Xin}\ \bibnamefont {Liu}}, \ and\
  \bibinfo {author} {\bibfnamefont {Xiao-Gang}\ \bibnamefont {Wen}},\
  }\bibfield  {title} {\enquote {\bibinfo {title} {Symmetry protected
  topological orders and the group cohomology of their symmetry group},}\
  }\href {\doibase 10.1103/PhysRevB.87.155114} {\bibfield  {journal} {\bibinfo
  {journal} {Phys.Rev.}\ }\textbf {\bibinfo {volume} {B87}},\ \bibinfo {pages}
  {155114} (\bibinfo {year} {2013}{\natexlab{a}})}\BibitemShut {NoStop}%
\bibitem [{\citenamefont {Senthil}(2015)}]{Senthil_2015}%
  \BibitemOpen
  \bibfield  {author} {\bibinfo {author} {\bibfnamefont {T.}~\bibnamefont
  {Senthil}},\ }\bibfield  {title} {\enquote {\bibinfo {title}
  {Symmetry-protected topological phases of quantum matter},}\ }\href {\doibase
  10.1146/annurev-conmatphys-031214-014740} {\bibfield  {journal} {\bibinfo
  {journal} {Annual Review of Condensed Matter Physics}\ }\textbf {\bibinfo
  {volume} {6}},\ \bibinfo {pages} {299–324} (\bibinfo {year}
  {2015})}\BibitemShut {NoStop}%
\bibitem [{\citenamefont {Pollmann}\ \emph {et~al.}(2010)\citenamefont
  {Pollmann}, \citenamefont {Berg}, \citenamefont {Turner},\ and\ \citenamefont
  {Oshikawa}}]{pollmann_entanglement_2010}%
  \BibitemOpen
  \bibfield  {author} {\bibinfo {author} {\bibfnamefont {Frank}\ \bibnamefont
  {Pollmann}}, \bibinfo {author} {\bibfnamefont {Erez}\ \bibnamefont {Berg}},
  \bibinfo {author} {\bibfnamefont {Ari~M.}\ \bibnamefont {Turner}}, \ and\
  \bibinfo {author} {\bibfnamefont {Masaki}\ \bibnamefont {Oshikawa}},\
  }\bibfield  {title} {\enquote {\bibinfo {title} {Entanglement spectrum of a
  topological phase in one dimension},}\ }\href {\doibase
  10.1103/PhysRevB.81.064439} {\bibfield  {journal} {\bibinfo  {journal}
  {Physical Review B}\ }\textbf {\bibinfo {volume} {81}} (\bibinfo {year}
  {2010}),\ 10.1103/PhysRevB.81.064439},\ \bibinfo {note} {arXiv:
  0910.1811}\BibitemShut {NoStop}%
\bibitem [{\citenamefont {Turner}\ \emph {et~al.}(2011)\citenamefont {Turner},
  \citenamefont {Pollmann},\ and\ \citenamefont {Berg}}]{Turner11}%
  \BibitemOpen
  \bibfield  {author} {\bibinfo {author} {\bibfnamefont {Ari~M.}\ \bibnamefont
  {Turner}}, \bibinfo {author} {\bibfnamefont {Frank}\ \bibnamefont
  {Pollmann}}, \ and\ \bibinfo {author} {\bibfnamefont {Erez}\ \bibnamefont
  {Berg}},\ }\bibfield  {title} {\enquote {\bibinfo {title} {Topological phases
  of one-dimensional fermions: An entanglement point of view},}\ }\href
  {\doibase 10.1103/PhysRevB.83.075102} {\bibfield  {journal} {\bibinfo
  {journal} {Phys. Rev. B}\ }\textbf {\bibinfo {volume} {83}},\ \bibinfo
  {pages} {075102} (\bibinfo {year} {2011})}\BibitemShut {NoStop}%
\bibitem [{\citenamefont {Fidkowski}\ and\ \citenamefont
  {Kitaev}(2011)}]{Fidkowski_2011}%
  \BibitemOpen
  \bibfield  {author} {\bibinfo {author} {\bibfnamefont {Lukasz}\ \bibnamefont
  {Fidkowski}}\ and\ \bibinfo {author} {\bibfnamefont {Alexei}\ \bibnamefont
  {Kitaev}},\ }\bibfield  {title} {\enquote {\bibinfo {title} {Topological
  phases of fermions in one dimension},}\ }\href {\doibase
  10.1103/physrevb.83.075103} {\bibfield  {journal} {\bibinfo  {journal}
  {Physical Review B}\ }\textbf {\bibinfo {volume} {83}} (\bibinfo {year}
  {2011}),\ 10.1103/physrevb.83.075103}\BibitemShut {NoStop}%
\bibitem [{\citenamefont {Chen}\ \emph {et~al.}(2011)\citenamefont {Chen},
  \citenamefont {Gu},\ and\ \citenamefont {Wen}}]{chen_complete_2011}%
  \BibitemOpen
  \bibfield  {author} {\bibinfo {author} {\bibfnamefont {Xie}\ \bibnamefont
  {Chen}}, \bibinfo {author} {\bibfnamefont {Zheng-Cheng}\ \bibnamefont {Gu}},
  \ and\ \bibinfo {author} {\bibfnamefont {Xiao-Gang}\ \bibnamefont {Wen}},\
  }\bibfield  {title} {\enquote {\bibinfo {title} {Complete classification of
  1d gapped quantum phases in interacting spin systems},}\ }\href {\doibase
  10.1103/PhysRevB.84.235128} {\bibfield  {journal} {\bibinfo  {journal}
  {Physical Review B}\ }\textbf {\bibinfo {volume} {84}} (\bibinfo {year}
  {2011}),\ 10.1103/PhysRevB.84.235128},\ \bibinfo {note} {arXiv:
  1103.3323}\BibitemShut {NoStop}%
\bibitem [{\citenamefont {{Schuch, N. and P\'erez-Garc\'ia, D. and Cirac, J.
  I.}}(2011)}]{Schuch11}%
  \BibitemOpen
  \bibfield  {author} {\bibinfo {author} {\bibnamefont {{Schuch, N. and
  P\'erez-Garc\'ia, D. and Cirac, J. I.}}},\ }\bibfield  {title} {\enquote
  {\bibinfo {title} {Classifying quantum phases using matrix product states and
  projected entangled pair states},}\ }\href {\doibase
  10.1103/PhysRevB.84.165139} {\bibfield  {journal} {\bibinfo  {journal} {Phys.
  Rev. B}\ }\textbf {\bibinfo {volume} {84}},\ \bibinfo {pages} {165139}
  (\bibinfo {year} {2011})}\BibitemShut {NoStop}%
\bibitem [{\citenamefont {Affleck}\ \emph {et~al.}(1988)\citenamefont
  {Affleck}, \citenamefont {Kennedy}, \citenamefont {Lieb},\ and\ \citenamefont
  {Tasaki}}]{affleck1988}%
  \BibitemOpen
  \bibfield  {author} {\bibinfo {author} {\bibfnamefont {Ian}\ \bibnamefont
  {Affleck}}, \bibinfo {author} {\bibfnamefont {Tom}\ \bibnamefont {Kennedy}},
  \bibinfo {author} {\bibfnamefont {Elliott~H.}\ \bibnamefont {Lieb}}, \ and\
  \bibinfo {author} {\bibfnamefont {Hal}\ \bibnamefont {Tasaki}},\ }\bibfield
  {title} {\enquote {\bibinfo {title} {Valence bond ground states in isotropic
  quantum antiferromagnets},}\ }\href
  {https://projecteuclid.org:443/euclid.cmp/1104161001} {\bibfield  {journal}
  {\bibinfo  {journal} {Comm. Math. Phys.}\ }\textbf {\bibinfo {volume}
  {115}},\ \bibinfo {pages} {477--528} (\bibinfo {year} {1988})}\BibitemShut
  {NoStop}%
\bibitem [{\citenamefont {Kennedy}(1990)}]{Kennedy90}%
  \BibitemOpen
  \bibfield  {author} {\bibinfo {author} {\bibfnamefont {T}~\bibnamefont
  {Kennedy}},\ }\bibfield  {title} {\enquote {\bibinfo {title} {Exact
  diagonalisations of open spin-1 chains},}\ }\href {\doibase
  10.1088/0953-8984/2/26/010} {\bibfield  {journal} {\bibinfo  {journal}
  {Journal of Physics: Condensed Matter}\ }\textbf {\bibinfo {volume} {2}},\
  \bibinfo {pages} {5737--5745} (\bibinfo {year} {1990})}\BibitemShut {NoStop}%
\bibitem [{\citenamefont {Kestner}\ \emph {et~al.}(2011)\citenamefont
  {Kestner}, \citenamefont {Wang}, \citenamefont {Sau},\ and\ \citenamefont
  {Das~Sarma}}]{Kestner11}%
  \BibitemOpen
  \bibfield  {author} {\bibinfo {author} {\bibfnamefont {J.~P.}\ \bibnamefont
  {Kestner}}, \bibinfo {author} {\bibfnamefont {Bin}\ \bibnamefont {Wang}},
  \bibinfo {author} {\bibfnamefont {Jay~D.}\ \bibnamefont {Sau}}, \ and\
  \bibinfo {author} {\bibfnamefont {S.}~\bibnamefont {Das~Sarma}},\ }\bibfield
  {title} {\enquote {\bibinfo {title} {{Prediction of a gapless topological
  Haldane liquid phase in a one-dimensional cold polar molecular lattice}},}\
  }\href {\doibase 10.1103/PhysRevB.83.174409} {\bibfield  {journal} {\bibinfo
  {journal} {Phys. Rev. B}\ }\textbf {\bibinfo {volume} {83}},\ \bibinfo
  {pages} {174409} (\bibinfo {year} {2011})}\BibitemShut {NoStop}%
\bibitem [{\citenamefont {Cheng}\ and\ \citenamefont {Tu}(2011)}]{Cheng11}%
  \BibitemOpen
  \bibfield  {author} {\bibinfo {author} {\bibfnamefont {Meng}\ \bibnamefont
  {Cheng}}\ and\ \bibinfo {author} {\bibfnamefont {Hong-Hao}\ \bibnamefont
  {Tu}},\ }\bibfield  {title} {\enquote {\bibinfo {title} {Majorana edge states
  in interacting two-chain ladders of fermions},}\ }\href {\doibase
  10.1103/PhysRevB.84.094503} {\bibfield  {journal} {\bibinfo  {journal} {Phys.
  Rev. B}\ }\textbf {\bibinfo {volume} {84}},\ \bibinfo {pages} {094503}
  (\bibinfo {year} {2011})}\BibitemShut {NoStop}%
\bibitem [{\citenamefont {Fidkowski}\ \emph {et~al.}(2011)\citenamefont
  {Fidkowski}, \citenamefont {Lutchyn}, \citenamefont {Nayak},\ and\
  \citenamefont {Fisher}}]{Fidkowski11longrange}%
  \BibitemOpen
  \bibfield  {author} {\bibinfo {author} {\bibfnamefont {Lukasz}\ \bibnamefont
  {Fidkowski}}, \bibinfo {author} {\bibfnamefont {Roman~M.}\ \bibnamefont
  {Lutchyn}}, \bibinfo {author} {\bibfnamefont {Chetan}\ \bibnamefont {Nayak}},
  \ and\ \bibinfo {author} {\bibfnamefont {Matthew P.~A.}\ \bibnamefont
  {Fisher}},\ }\bibfield  {title} {\enquote {\bibinfo {title} {Majorana zero
  modes in one-dimensional quantum wires without long-ranged superconducting
  order},}\ }\href {\doibase 10.1103/PhysRevB.84.195436} {\bibfield  {journal}
  {\bibinfo  {journal} {Phys. Rev. B}\ }\textbf {\bibinfo {volume} {84}},\
  \bibinfo {pages} {195436} (\bibinfo {year} {2011})}\BibitemShut {NoStop}%
\bibitem [{\citenamefont {Sau}\ \emph {et~al.}(2011)\citenamefont {Sau},
  \citenamefont {Halperin}, \citenamefont {Flensberg},\ and\ \citenamefont
  {Das~Sarma}}]{Sau11}%
  \BibitemOpen
  \bibfield  {author} {\bibinfo {author} {\bibfnamefont {Jay~D.}\ \bibnamefont
  {Sau}}, \bibinfo {author} {\bibfnamefont {B.~I.}\ \bibnamefont {Halperin}},
  \bibinfo {author} {\bibfnamefont {K.}~\bibnamefont {Flensberg}}, \ and\
  \bibinfo {author} {\bibfnamefont {S.}~\bibnamefont {Das~Sarma}},\ }\bibfield
  {title} {\enquote {\bibinfo {title} {Number conserving theory for
  topologically protected degeneracy in one-dimensional fermions},}\ }\href
  {\doibase 10.1103/PhysRevB.84.144509} {\bibfield  {journal} {\bibinfo
  {journal} {Phys. Rev. B}\ }\textbf {\bibinfo {volume} {84}},\ \bibinfo
  {pages} {144509} (\bibinfo {year} {2011})}\BibitemShut {NoStop}%
\bibitem [{\citenamefont {Ruhman}\ \emph {et~al.}(2012)\citenamefont {Ruhman},
  \citenamefont {Dalla~Torre}, \citenamefont {Huber},\ and\ \citenamefont
  {Altman}}]{Ruhman12}%
  \BibitemOpen
  \bibfield  {author} {\bibinfo {author} {\bibfnamefont {J.}~\bibnamefont
  {Ruhman}}, \bibinfo {author} {\bibfnamefont {E.~G.}\ \bibnamefont
  {Dalla~Torre}}, \bibinfo {author} {\bibfnamefont {S.~D.}\ \bibnamefont
  {Huber}}, \ and\ \bibinfo {author} {\bibfnamefont {E.}~\bibnamefont
  {Altman}},\ }\bibfield  {title} {\enquote {\bibinfo {title} {Nonlocal order
  in elongated dipolar gases},}\ }\href {\doibase 10.1103/PhysRevB.85.125121}
  {\bibfield  {journal} {\bibinfo  {journal} {Phys. Rev. B}\ }\textbf {\bibinfo
  {volume} {85}},\ \bibinfo {pages} {125121} (\bibinfo {year}
  {2012})}\BibitemShut {NoStop}%
\bibitem [{\citenamefont {{Grover}}\ and\ \citenamefont
  {{Vishwanath}}(2012)}]{Grover12}%
  \BibitemOpen
  \bibfield  {author} {\bibinfo {author} {\bibfnamefont {Tarun}\ \bibnamefont
  {{Grover}}}\ and\ \bibinfo {author} {\bibfnamefont {Ashvin}\ \bibnamefont
  {{Vishwanath}}},\ }\bibfield  {title} {\enquote {\bibinfo {title} {{Quantum
  Criticality in Topological Insulators and Superconductors: Emergence of
  Strongly Coupled Majoranas and Supersymmetry}},}\ }\href@noop {} {\bibfield
  {journal} {\bibinfo  {journal} {arXiv e-prints}\ ,\ \bibinfo {eid}
  {arXiv:1206.1332}} (\bibinfo {year} {2012})},\ \Eprint
  {http://arxiv.org/abs/1206.1332} {arXiv:1206.1332 [cond-mat.str-el]}
  \BibitemShut {NoStop}%
\bibitem [{\citenamefont {Kraus}\ \emph {et~al.}(2013)\citenamefont {Kraus},
  \citenamefont {Dalmonte}, \citenamefont {Baranov}, \citenamefont
  {L\"auchli},\ and\ \citenamefont {Zoller}}]{Kraus13}%
  \BibitemOpen
  \bibfield  {author} {\bibinfo {author} {\bibfnamefont {Christina~V.}\
  \bibnamefont {Kraus}}, \bibinfo {author} {\bibfnamefont {Marcello}\
  \bibnamefont {Dalmonte}}, \bibinfo {author} {\bibfnamefont {Mikhail~A.}\
  \bibnamefont {Baranov}}, \bibinfo {author} {\bibfnamefont {Andreas~M.}\
  \bibnamefont {L\"auchli}}, \ and\ \bibinfo {author} {\bibfnamefont
  {P.}~\bibnamefont {Zoller}},\ }\bibfield  {title} {\enquote {\bibinfo {title}
  {{Majorana Edge States in Atomic Wires Coupled by Pair Hopping}},}\ }\href
  {\doibase 10.1103/PhysRevLett.111.173004} {\bibfield  {journal} {\bibinfo
  {journal} {Phys. Rev. Lett.}\ }\textbf {\bibinfo {volume} {111}},\ \bibinfo
  {pages} {173004} (\bibinfo {year} {2013})}\BibitemShut {NoStop}%
\bibitem [{\citenamefont {Ortiz}\ \emph {et~al.}(2014)\citenamefont {Ortiz},
  \citenamefont {Dukelsky}, \citenamefont {Cobanera}, \citenamefont {Esebbag},\
  and\ \citenamefont {Beenakker}}]{Ortiz14}%
  \BibitemOpen
  \bibfield  {author} {\bibinfo {author} {\bibfnamefont {Gerardo}\ \bibnamefont
  {Ortiz}}, \bibinfo {author} {\bibfnamefont {Jorge}\ \bibnamefont {Dukelsky}},
  \bibinfo {author} {\bibfnamefont {Emilio}\ \bibnamefont {Cobanera}}, \bibinfo
  {author} {\bibfnamefont {Carlos}\ \bibnamefont {Esebbag}}, \ and\ \bibinfo
  {author} {\bibfnamefont {Carlo}\ \bibnamefont {Beenakker}},\ }\bibfield
  {title} {\enquote {\bibinfo {title} {{Many-Body Characterization of
  Particle-Conserving Topological Superfluids}},}\ }\href {\doibase
  10.1103/PhysRevLett.113.267002} {\bibfield  {journal} {\bibinfo  {journal}
  {Phys. Rev. Lett.}\ }\textbf {\bibinfo {volume} {113}},\ \bibinfo {pages}
  {267002} (\bibinfo {year} {2014})}\BibitemShut {NoStop}%
\bibitem [{\citenamefont {Keselman}\ and\ \citenamefont
  {Berg}(2015)}]{Keselman15}%
  \BibitemOpen
  \bibfield  {author} {\bibinfo {author} {\bibfnamefont {Anna}\ \bibnamefont
  {Keselman}}\ and\ \bibinfo {author} {\bibfnamefont {Erez}\ \bibnamefont
  {Berg}},\ }\bibfield  {title} {\enquote {\bibinfo {title} {Gapless
  symmetry-protected topological phase of fermions in one dimension},}\ }\href
  {\doibase 10.1103/PhysRevB.91.235309} {\bibfield  {journal} {\bibinfo
  {journal} {Phys. Rev. B}\ }\textbf {\bibinfo {volume} {91}},\ \bibinfo
  {pages} {235309} (\bibinfo {year} {2015})}\BibitemShut {NoStop}%
\bibitem [{\citenamefont {Ruhman}\ \emph {et~al.}(2015)\citenamefont {Ruhman},
  \citenamefont {Berg},\ and\ \citenamefont {Altman}}]{Ruhman15}%
  \BibitemOpen
  \bibfield  {author} {\bibinfo {author} {\bibfnamefont {Jonathan}\
  \bibnamefont {Ruhman}}, \bibinfo {author} {\bibfnamefont {Erez}\ \bibnamefont
  {Berg}}, \ and\ \bibinfo {author} {\bibfnamefont {Ehud}\ \bibnamefont
  {Altman}},\ }\bibfield  {title} {\enquote {\bibinfo {title} {{Topological
  States in a One-Dimensional Fermi Gas with Attractive Interaction}},}\ }\href
  {\doibase 10.1103/PhysRevLett.114.100401} {\bibfield  {journal} {\bibinfo
  {journal} {Phys. Rev. Lett.}\ }\textbf {\bibinfo {volume} {114}},\ \bibinfo
  {pages} {100401} (\bibinfo {year} {2015})}\BibitemShut {NoStop}%
\bibitem [{\citenamefont {Kainaris}\ and\ \citenamefont
  {Carr}(2015)}]{Kainaris15}%
  \BibitemOpen
  \bibfield  {author} {\bibinfo {author} {\bibfnamefont {Nikolaos}\
  \bibnamefont {Kainaris}}\ and\ \bibinfo {author} {\bibfnamefont {Sam~T.}\
  \bibnamefont {Carr}},\ }\bibfield  {title} {\enquote {\bibinfo {title}
  {Emergent topological properties in interacting one-dimensional systems with
  spin-orbit coupling},}\ }\href {\doibase 10.1103/PhysRevB.92.035139}
  {\bibfield  {journal} {\bibinfo  {journal} {Phys. Rev. B}\ }\textbf {\bibinfo
  {volume} {92}},\ \bibinfo {pages} {035139} (\bibinfo {year}
  {2015})}\BibitemShut {NoStop}%
\bibitem [{\citenamefont {Iemini}\ \emph {et~al.}(2015)\citenamefont {Iemini},
  \citenamefont {Mazza}, \citenamefont {Rossini}, \citenamefont {Fazio},\ and\
  \citenamefont {Diehl}}]{Iemini15}%
  \BibitemOpen
  \bibfield  {author} {\bibinfo {author} {\bibfnamefont {Fernando}\
  \bibnamefont {Iemini}}, \bibinfo {author} {\bibfnamefont {Leonardo}\
  \bibnamefont {Mazza}}, \bibinfo {author} {\bibfnamefont {Davide}\
  \bibnamefont {Rossini}}, \bibinfo {author} {\bibfnamefont {Rosario}\
  \bibnamefont {Fazio}}, \ and\ \bibinfo {author} {\bibfnamefont {Sebastian}\
  \bibnamefont {Diehl}},\ }\bibfield  {title} {\enquote {\bibinfo {title}
  {{Localized Majorana-Like Modes in a Number-Conserving Setting: An Exactly
  Solvable Model}},}\ }\href {\doibase 10.1103/PhysRevLett.115.156402}
  {\bibfield  {journal} {\bibinfo  {journal} {Phys. Rev. Lett.}\ }\textbf
  {\bibinfo {volume} {115}},\ \bibinfo {pages} {156402} (\bibinfo {year}
  {2015})}\BibitemShut {NoStop}%
\bibitem [{\citenamefont {Lang}\ and\ \citenamefont
  {B\"uchler}(2015)}]{Lang15}%
  \BibitemOpen
  \bibfield  {author} {\bibinfo {author} {\bibfnamefont {Nicolai}\ \bibnamefont
  {Lang}}\ and\ \bibinfo {author} {\bibfnamefont {Hans~Peter}\ \bibnamefont
  {B\"uchler}},\ }\bibfield  {title} {\enquote {\bibinfo {title} {Topological
  states in a microscopic model of interacting fermions},}\ }\href {\doibase
  10.1103/PhysRevB.92.041118} {\bibfield  {journal} {\bibinfo  {journal} {Phys.
  Rev. B}\ }\textbf {\bibinfo {volume} {92}},\ \bibinfo {pages} {041118(R)}
  (\bibinfo {year} {2015})}\BibitemShut {NoStop}%
\bibitem [{\citenamefont {Ortiz}\ and\ \citenamefont
  {Cobanera}(2016)}]{Ortiz16}%
  \BibitemOpen
  \bibfield  {author} {\bibinfo {author} {\bibfnamefont {Gerardo}\ \bibnamefont
  {Ortiz}}\ and\ \bibinfo {author} {\bibfnamefont {Emilio}\ \bibnamefont
  {Cobanera}},\ }\bibfield  {title} {\enquote {\bibinfo {title} {{What is a
  particle-conserving Topological Superfluid? The fate of Majorana modes beyond
  mean-field theory}},}\ }\href {\doibase
  https://doi.org/10.1016/j.aop.2016.05.020} {\bibfield  {journal} {\bibinfo
  {journal} {Annals of Physics}\ }\textbf {\bibinfo {volume} {372}},\ \bibinfo
  {pages} {357 -- 374} (\bibinfo {year} {2016})}\BibitemShut {NoStop}%
\bibitem [{\citenamefont {Montorsi}\ \emph {et~al.}(2017)\citenamefont
  {Montorsi}, \citenamefont {Dolcini}, \citenamefont {Iotti},\ and\
  \citenamefont {Rossi}}]{Montorsi17}%
  \BibitemOpen
  \bibfield  {author} {\bibinfo {author} {\bibfnamefont {Arianna}\ \bibnamefont
  {Montorsi}}, \bibinfo {author} {\bibfnamefont {Fabrizio}\ \bibnamefont
  {Dolcini}}, \bibinfo {author} {\bibfnamefont {Rita~C.}\ \bibnamefont
  {Iotti}}, \ and\ \bibinfo {author} {\bibfnamefont {Fausto}\ \bibnamefont
  {Rossi}},\ }\bibfield  {title} {\enquote {\bibinfo {title}
  {Symmetry-protected topological phases of one-dimensional interacting
  fermions with spin-charge separation},}\ }\href {\doibase
  10.1103/PhysRevB.95.245108} {\bibfield  {journal} {\bibinfo  {journal} {Phys.
  Rev. B}\ }\textbf {\bibinfo {volume} {95}},\ \bibinfo {pages} {245108}
  (\bibinfo {year} {2017})}\BibitemShut {NoStop}%
\bibitem [{\citenamefont {Wang}\ \emph
  {et~al.}(2017{\natexlab{a}})\citenamefont {Wang}, \citenamefont {Xu},
  \citenamefont {Pu},\ and\ \citenamefont {Hazzard}}]{Wang17}%
  \BibitemOpen
  \bibfield  {author} {\bibinfo {author} {\bibfnamefont {Zhiyuan}\ \bibnamefont
  {Wang}}, \bibinfo {author} {\bibfnamefont {Youjiang}\ \bibnamefont {Xu}},
  \bibinfo {author} {\bibfnamefont {Han}\ \bibnamefont {Pu}}, \ and\ \bibinfo
  {author} {\bibfnamefont {Kaden R.~A.}\ \bibnamefont {Hazzard}},\ }\bibfield
  {title} {\enquote {\bibinfo {title} {Number-conserving interacting fermion
  models with exact topological superconducting ground states},}\ }\href
  {\doibase 10.1103/PhysRevB.96.115110} {\bibfield  {journal} {\bibinfo
  {journal} {Phys. Rev. B}\ }\textbf {\bibinfo {volume} {96}},\ \bibinfo
  {pages} {115110} (\bibinfo {year} {2017}{\natexlab{a}})}\BibitemShut
  {NoStop}%
\bibitem [{\citenamefont {Ruhman}\ and\ \citenamefont
  {Altman}(2017)}]{Ruhman17}%
  \BibitemOpen
  \bibfield  {author} {\bibinfo {author} {\bibfnamefont {Jonathan}\
  \bibnamefont {Ruhman}}\ and\ \bibinfo {author} {\bibfnamefont {Ehud}\
  \bibnamefont {Altman}},\ }\bibfield  {title} {\enquote {\bibinfo {title}
  {Topological degeneracy and pairing in a one-dimensional gas of spinless
  fermions},}\ }\href {\doibase 10.1103/PhysRevB.96.085133} {\bibfield
  {journal} {\bibinfo  {journal} {Phys. Rev. B}\ }\textbf {\bibinfo {volume}
  {96}},\ \bibinfo {pages} {085133} (\bibinfo {year} {2017})}\BibitemShut
  {NoStop}%
\bibitem [{\citenamefont {Scaffidi}\ \emph {et~al.}(2017)\citenamefont
  {Scaffidi}, \citenamefont {Parker},\ and\ \citenamefont
  {Vasseur}}]{Scaffidi17}%
  \BibitemOpen
  \bibfield  {author} {\bibinfo {author} {\bibfnamefont {Thomas}\ \bibnamefont
  {Scaffidi}}, \bibinfo {author} {\bibfnamefont {Daniel~E.}\ \bibnamefont
  {Parker}}, \ and\ \bibinfo {author} {\bibfnamefont {Romain}\ \bibnamefont
  {Vasseur}},\ }\bibfield  {title} {\enquote {\bibinfo {title} {{Gapless
  Symmetry-Protected Topological Order}},}\ }\href {\doibase
  10.1103/PhysRevX.7.041048} {\bibfield  {journal} {\bibinfo  {journal} {Phys.
  Rev. X}\ }\textbf {\bibinfo {volume} {7}},\ \bibinfo {pages} {041048}
  (\bibinfo {year} {2017})}\BibitemShut {NoStop}%
\bibitem [{\citenamefont {Guther}\ \emph {et~al.}(2017)\citenamefont {Guther},
  \citenamefont {Lang},\ and\ \citenamefont {B\"uchler}}]{Guther17}%
  \BibitemOpen
  \bibfield  {author} {\bibinfo {author} {\bibfnamefont {K.}~\bibnamefont
  {Guther}}, \bibinfo {author} {\bibfnamefont {N.}~\bibnamefont {Lang}}, \ and\
  \bibinfo {author} {\bibfnamefont {H.~P.}\ \bibnamefont {B\"uchler}},\
  }\bibfield  {title} {\enquote {\bibinfo {title} {{Ising anyonic topological
  phase of interacting fermions in one dimension}},}\ }\href {\doibase
  10.1103/PhysRevB.96.121109} {\bibfield  {journal} {\bibinfo  {journal} {Phys.
  Rev. B}\ }\textbf {\bibinfo {volume} {96}},\ \bibinfo {pages} {121109}
  (\bibinfo {year} {2017})}\BibitemShut {NoStop}%
\bibitem [{\citenamefont {Kainaris}\ \emph {et~al.}(2017)\citenamefont
  {Kainaris}, \citenamefont {Santos}, \citenamefont {Gutman},\ and\
  \citenamefont {Carr}}]{Kainaris17}%
  \BibitemOpen
  \bibfield  {author} {\bibinfo {author} {\bibfnamefont {Nikolaos}\
  \bibnamefont {Kainaris}}, \bibinfo {author} {\bibfnamefont {Raul~A.}\
  \bibnamefont {Santos}}, \bibinfo {author} {\bibfnamefont {D.~B.}\
  \bibnamefont {Gutman}}, \ and\ \bibinfo {author} {\bibfnamefont {Sam~T.}\
  \bibnamefont {Carr}},\ }\bibfield  {title} {\enquote {\bibinfo {title}
  {Interaction induced topological protection in one-dimensional conductors},}\
  }\href {\doibase 10.1002/prop.201600054} {\bibfield  {journal} {\bibinfo
  {journal} {Fortschritte der Physik}\ }\textbf {\bibinfo {volume} {65}},\
  \bibinfo {pages} {1600054} (\bibinfo {year} {2017})}\BibitemShut {NoStop}%
\bibitem [{\citenamefont {Jiang}\ \emph {et~al.}(2018)\citenamefont {Jiang},
  \citenamefont {Li}, \citenamefont {Seidel},\ and\ \citenamefont
  {Lee}}]{Jiang18}%
  \BibitemOpen
  \bibfield  {author} {\bibinfo {author} {\bibfnamefont {Hong-Chen}\
  \bibnamefont {Jiang}}, \bibinfo {author} {\bibfnamefont {Zi-Xiang}\
  \bibnamefont {Li}}, \bibinfo {author} {\bibfnamefont {Alexander}\
  \bibnamefont {Seidel}}, \ and\ \bibinfo {author} {\bibfnamefont {Dung-Hai}\
  \bibnamefont {Lee}},\ }\bibfield  {title} {\enquote {\bibinfo {title}
  {{Symmetry protected topological Luttinger liquids and the phase transition
  between them}},}\ }\href {\doibase
  https://doi.org/10.1016/j.scib.2018.05.010} {\bibfield  {journal} {\bibinfo
  {journal} {Science Bulletin}\ }\textbf {\bibinfo {volume} {63}},\ \bibinfo
  {pages} {753 -- 758} (\bibinfo {year} {2018})}\BibitemShut {NoStop}%
\bibitem [{\citenamefont {Zhang}\ and\ \citenamefont {Liu}(2018)}]{Zhang18}%
  \BibitemOpen
  \bibfield  {author} {\bibinfo {author} {\bibfnamefont {Rui-Xing}\
  \bibnamefont {Zhang}}\ and\ \bibinfo {author} {\bibfnamefont {Chao-Xing}\
  \bibnamefont {Liu}},\ }\bibfield  {title} {\enquote {\bibinfo {title}
  {{Crystalline Symmetry-Protected Majorana Mode in Number-Conserving Dirac
  Semimetal Nanowires}},}\ }\href {\doibase 10.1103/PhysRevLett.120.156802}
  {\bibfield  {journal} {\bibinfo  {journal} {Phys. Rev. Lett.}\ }\textbf
  {\bibinfo {volume} {120}},\ \bibinfo {pages} {156802} (\bibinfo {year}
  {2018})}\BibitemShut {NoStop}%
\bibitem [{\citenamefont {Verresen}\ \emph {et~al.}(2018)\citenamefont
  {Verresen}, \citenamefont {Jones},\ and\ \citenamefont
  {Pollmann}}]{Verresen18}%
  \BibitemOpen
  \bibfield  {author} {\bibinfo {author} {\bibfnamefont {Ruben}\ \bibnamefont
  {Verresen}}, \bibinfo {author} {\bibfnamefont {Nick~G.}\ \bibnamefont
  {Jones}}, \ and\ \bibinfo {author} {\bibfnamefont {Frank}\ \bibnamefont
  {Pollmann}},\ }\bibfield  {title} {\enquote {\bibinfo {title} {{Topology and
  Edge Modes in Quantum Critical Chains}},}\ }\href {\doibase
  10.1103/PhysRevLett.120.057001} {\bibfield  {journal} {\bibinfo  {journal}
  {Phys. Rev. Lett.}\ }\textbf {\bibinfo {volume} {120}},\ \bibinfo {pages}
  {057001} (\bibinfo {year} {2018})}\BibitemShut {NoStop}%
\bibitem [{\citenamefont {Parker}\ \emph {et~al.}(2018)\citenamefont {Parker},
  \citenamefont {Scaffidi},\ and\ \citenamefont {Vasseur}}]{Parker18}%
  \BibitemOpen
  \bibfield  {author} {\bibinfo {author} {\bibfnamefont {Daniel~E.}\
  \bibnamefont {Parker}}, \bibinfo {author} {\bibfnamefont {Thomas}\
  \bibnamefont {Scaffidi}}, \ and\ \bibinfo {author} {\bibfnamefont {Romain}\
  \bibnamefont {Vasseur}},\ }\bibfield  {title} {\enquote {\bibinfo {title}
  {{Topological Luttinger liquids from decorated domain walls}},}\ }\href
  {\doibase 10.1103/PhysRevB.97.165114} {\bibfield  {journal} {\bibinfo
  {journal} {Phys. Rev. B}\ }\textbf {\bibinfo {volume} {97}},\ \bibinfo
  {pages} {165114} (\bibinfo {year} {2018})}\BibitemShut {NoStop}%
\bibitem [{\citenamefont {Keselman}\ \emph {et~al.}(2018)\citenamefont
  {Keselman}, \citenamefont {Berg},\ and\ \citenamefont {Azaria}}]{Keselman18}%
  \BibitemOpen
  \bibfield  {author} {\bibinfo {author} {\bibfnamefont {Anna}\ \bibnamefont
  {Keselman}}, \bibinfo {author} {\bibfnamefont {Erez}\ \bibnamefont {Berg}}, \
  and\ \bibinfo {author} {\bibfnamefont {Patrick}\ \bibnamefont {Azaria}},\
  }\bibfield  {title} {\enquote {\bibinfo {title} {{From one-dimensional charge
  conserving superconductors to the gapless Haldane phase}},}\ }\href {\doibase
  10.1103/PhysRevB.98.214501} {\bibfield  {journal} {\bibinfo  {journal} {Phys.
  Rev. B}\ }\textbf {\bibinfo {volume} {98}},\ \bibinfo {pages} {214501}
  (\bibinfo {year} {2018})}\BibitemShut {NoStop}%
\bibitem [{\citenamefont {Chen}\ \emph {et~al.}(2018)\citenamefont {Chen},
  \citenamefont {Yan}, \citenamefont {Ting}, \citenamefont {Chen},\ and\
  \citenamefont {Burnell}}]{Chen18}%
  \BibitemOpen
  \bibfield  {author} {\bibinfo {author} {\bibfnamefont {Chun}\ \bibnamefont
  {Chen}}, \bibinfo {author} {\bibfnamefont {Wei}\ \bibnamefont {Yan}},
  \bibinfo {author} {\bibfnamefont {C.~S.}\ \bibnamefont {Ting}}, \bibinfo
  {author} {\bibfnamefont {Yan}\ \bibnamefont {Chen}}, \ and\ \bibinfo {author}
  {\bibfnamefont {F.~J.}\ \bibnamefont {Burnell}},\ }\bibfield  {title}
  {\enquote {\bibinfo {title} {Flux-stabilized majorana zero modes in coupled
  one-dimensional fermi wires},}\ }\href {\doibase 10.1103/PhysRevB.98.161106}
  {\bibfield  {journal} {\bibinfo  {journal} {Phys. Rev. B}\ }\textbf {\bibinfo
  {volume} {98}},\ \bibinfo {pages} {161106} (\bibinfo {year}
  {2018})}\BibitemShut {NoStop}%
\bibitem [{\citenamefont {Verresen}\ \emph {et~al.}(2021)\citenamefont
  {Verresen}, \citenamefont {Thorngren}, \citenamefont {Jones},\ and\
  \citenamefont {Pollmann}}]{Verresen21gapless}%
  \BibitemOpen
  \bibfield  {author} {\bibinfo {author} {\bibfnamefont {Ruben}\ \bibnamefont
  {Verresen}}, \bibinfo {author} {\bibfnamefont {Ryan}\ \bibnamefont
  {Thorngren}}, \bibinfo {author} {\bibfnamefont {Nick~G.}\ \bibnamefont
  {Jones}}, \ and\ \bibinfo {author} {\bibfnamefont {Frank}\ \bibnamefont
  {Pollmann}},\ }\bibfield  {title} {\enquote {\bibinfo {title} {Gapless
  topological phases and symmetry-enriched quantum criticality},}\ }\href
  {\doibase 10.1103/PhysRevX.11.041059} {\bibfield  {journal} {\bibinfo
  {journal} {Phys. Rev. X}\ }\textbf {\bibinfo {volume} {11}},\ \bibinfo
  {pages} {041059} (\bibinfo {year} {2021})}\BibitemShut {NoStop}%
\bibitem [{\citenamefont {Duque}\ \emph {et~al.}(2021)\citenamefont {Duque},
  \citenamefont {Hu}, \citenamefont {You}, \citenamefont {Khemani},
  \citenamefont {Verresen},\ and\ \citenamefont {Vasseur}}]{Duque21}%
  \BibitemOpen
  \bibfield  {author} {\bibinfo {author} {\bibfnamefont {Carlos~M.}\
  \bibnamefont {Duque}}, \bibinfo {author} {\bibfnamefont {Hong-Ye}\
  \bibnamefont {Hu}}, \bibinfo {author} {\bibfnamefont {Yi-Zhuang}\
  \bibnamefont {You}}, \bibinfo {author} {\bibfnamefont {Vedika}\ \bibnamefont
  {Khemani}}, \bibinfo {author} {\bibfnamefont {Ruben}\ \bibnamefont
  {Verresen}}, \ and\ \bibinfo {author} {\bibfnamefont {Romain}\ \bibnamefont
  {Vasseur}},\ }\bibfield  {title} {\enquote {\bibinfo {title} {Topological and
  symmetry-enriched random quantum critical points},}\ }\href {\doibase
  10.1103/PhysRevB.103.L100207} {\bibfield  {journal} {\bibinfo  {journal}
  {Phys. Rev. B}\ }\textbf {\bibinfo {volume} {103}},\ \bibinfo {pages}
  {L100207} (\bibinfo {year} {2021})}\BibitemShut {NoStop}%
\bibitem [{\citenamefont {Balabanov}\ \emph {et~al.}(2021)\citenamefont
  {Balabanov}, \citenamefont {Erkensten},\ and\ \citenamefont
  {Johannesson}}]{Balabanov21}%
  \BibitemOpen
  \bibfield  {author} {\bibinfo {author} {\bibfnamefont {Oleksandr}\
  \bibnamefont {Balabanov}}, \bibinfo {author} {\bibfnamefont {Daniel}\
  \bibnamefont {Erkensten}}, \ and\ \bibinfo {author} {\bibfnamefont {Henrik}\
  \bibnamefont {Johannesson}},\ }\bibfield  {title} {\enquote {\bibinfo {title}
  {Topology of critical chiral phases: Multiband insulators and
  superconductors},}\ }\href {\doibase 10.1103/PhysRevResearch.3.043048}
  {\bibfield  {journal} {\bibinfo  {journal} {Phys. Rev. Research}\ }\textbf
  {\bibinfo {volume} {3}},\ \bibinfo {pages} {043048} (\bibinfo {year}
  {2021})}\BibitemShut {NoStop}%
\bibitem [{\citenamefont {{Chang}}\ and\ \citenamefont
  {{Hosur}}(2022)}]{Chang22}%
  \BibitemOpen
  \bibfield  {author} {\bibinfo {author} {\bibfnamefont {Shun-Chiao}\
  \bibnamefont {{Chang}}}\ and\ \bibinfo {author} {\bibfnamefont {Pavan}\
  \bibnamefont {{Hosur}}},\ }\href@noop {} {\enquote {\bibinfo {title}
  {{Absence of Friedel oscillations in the entanglement entropy profile of
  one-dimensional intrinsically gapless topological phases}},}\ } (\bibinfo
  {year} {2022}),\ \Eprint {http://arxiv.org/abs/2201.07260} {arXiv:2201.07260
  [cond-mat.str-el]} \BibitemShut {NoStop}%
\bibitem [{\citenamefont {Fraxanet}\ \emph {et~al.}(2022)\citenamefont
  {Fraxanet}, \citenamefont {Gonz\'alez-Cuadra}, \citenamefont {Pfau},
  \citenamefont {Lewenstein}, \citenamefont {Langen},\ and\ \citenamefont
  {Barbiero}}]{Fraxanet22}%
  \BibitemOpen
  \bibfield  {author} {\bibinfo {author} {\bibfnamefont {Joana}\ \bibnamefont
  {Fraxanet}}, \bibinfo {author} {\bibfnamefont {Daniel}\ \bibnamefont
  {Gonz\'alez-Cuadra}}, \bibinfo {author} {\bibfnamefont {Tilman}\ \bibnamefont
  {Pfau}}, \bibinfo {author} {\bibfnamefont {Maciej}\ \bibnamefont
  {Lewenstein}}, \bibinfo {author} {\bibfnamefont {Tim}\ \bibnamefont
  {Langen}}, \ and\ \bibinfo {author} {\bibfnamefont {Luca}\ \bibnamefont
  {Barbiero}},\ }\bibfield  {title} {\enquote {\bibinfo {title} {Topological
  quantum critical points in the extended bose-hubbard model},}\ }\href
  {\doibase 10.1103/PhysRevLett.128.043402} {\bibfield  {journal} {\bibinfo
  {journal} {Phys. Rev. Lett.}\ }\textbf {\bibinfo {volume} {128}},\ \bibinfo
  {pages} {043402} (\bibinfo {year} {2022})}\BibitemShut {NoStop}%
\bibitem [{\citenamefont {Balabanov}\ \emph {et~al.}(2022)\citenamefont
  {Balabanov}, \citenamefont {Ortega-Taberner},\ and\ \citenamefont
  {Hermanns}}]{Balabanov22}%
  \BibitemOpen
  \bibfield  {author} {\bibinfo {author} {\bibfnamefont {Oleksandr}\
  \bibnamefont {Balabanov}}, \bibinfo {author} {\bibfnamefont {Carlos}\
  \bibnamefont {Ortega-Taberner}}, \ and\ \bibinfo {author} {\bibfnamefont
  {Maria}\ \bibnamefont {Hermanns}},\ }\bibfield  {title} {\enquote {\bibinfo
  {title} {Quantization of topological indices in critical chains at low
  temperatures},}\ }\href {\doibase 10.1103/PhysRevB.106.045116} {\bibfield
  {journal} {\bibinfo  {journal} {Phys. Rev. B}\ }\textbf {\bibinfo {volume}
  {106}},\ \bibinfo {pages} {045116} (\bibinfo {year} {2022})}\BibitemShut
  {NoStop}%
\bibitem [{\citenamefont {{Tsui}}\ \emph {et~al.}(2015)\citenamefont {{Tsui}},
  \citenamefont {{Jiang}}, \citenamefont {{Lu}},\ and\ \citenamefont
  {{Lee}}}]{Tsui15}%
  \BibitemOpen
  \bibfield  {author} {\bibinfo {author} {\bibfnamefont {Lokman}\ \bibnamefont
  {{Tsui}}}, \bibinfo {author} {\bibfnamefont {Hong-Chen}\ \bibnamefont
  {{Jiang}}}, \bibinfo {author} {\bibfnamefont {Yuan-Ming}\ \bibnamefont
  {{Lu}}}, \ and\ \bibinfo {author} {\bibfnamefont {Dung-Hai}\ \bibnamefont
  {{Lee}}},\ }\bibfield  {title} {\enquote {\bibinfo {title} {{Quantum phase
  transitions between a class of symmetry protected topological states}},}\
  }\href {\doibase 10.1016/j.nuclphysb.2015.04.020} {\bibfield  {journal}
  {\bibinfo  {journal} {Nuclear Physics B}\ }\textbf {\bibinfo {volume}
  {896}},\ \bibinfo {pages} {330--359} (\bibinfo {year} {2015})},\ \Eprint
  {http://arxiv.org/abs/1503.06794} {arXiv:1503.06794 [cond-mat.str-el]}
  \BibitemShut {NoStop}%
\bibitem [{\citenamefont {Tsui}\ \emph {et~al.}(2017)\citenamefont {Tsui},
  \citenamefont {Huang}, \citenamefont {Jiang},\ and\ \citenamefont
  {Lee}}]{Tsui17}%
  \BibitemOpen
  \bibfield  {author} {\bibinfo {author} {\bibfnamefont {Lokman}\ \bibnamefont
  {Tsui}}, \bibinfo {author} {\bibfnamefont {Yen-Ta}\ \bibnamefont {Huang}},
  \bibinfo {author} {\bibfnamefont {Hong-Chen}\ \bibnamefont {Jiang}}, \ and\
  \bibinfo {author} {\bibfnamefont {Dung-Hai}\ \bibnamefont {Lee}},\ }\bibfield
   {title} {\enquote {\bibinfo {title} {{The phase transitions between $Z_n
  \times Z_n$ bosonic topological phases in 1+1D, and a constraint on the
  central charge for the critical points between bosonic symmetry protected
  topological phases}},}\ }\href {\doibase
  https://doi.org/10.1016/j.nuclphysb.2017.03.021} {\bibfield  {journal}
  {\bibinfo  {journal} {Nuclear Physics B}\ }\textbf {\bibinfo {volume}
  {919}},\ \bibinfo {pages} {470 -- 503} (\bibinfo {year} {2017})}\BibitemShut
  {NoStop}%
\bibitem [{\citenamefont {Bultinck}(2019)}]{Bultinck19}%
  \BibitemOpen
  \bibfield  {author} {\bibinfo {author} {\bibfnamefont {Nick}\ \bibnamefont
  {Bultinck}},\ }\bibfield  {title} {\enquote {\bibinfo {title} {Uv perspective
  on mixed anomalies at critical points between bosonic symmetry-protected
  phases},}\ }\href {\doibase 10.1103/PhysRevB.100.165132} {\bibfield
  {journal} {\bibinfo  {journal} {Phys. Rev. B}\ }\textbf {\bibinfo {volume}
  {100}},\ \bibinfo {pages} {165132} (\bibinfo {year} {2019})}\BibitemShut
  {NoStop}%
\bibitem [{\citenamefont {Dupont}\ \emph
  {et~al.}(2021{\natexlab{a}})\citenamefont {Dupont}, \citenamefont {Gazit},\
  and\ \citenamefont {Scaffidi}}]{Dupont21}%
  \BibitemOpen
  \bibfield  {author} {\bibinfo {author} {\bibfnamefont {Maxime}\ \bibnamefont
  {Dupont}}, \bibinfo {author} {\bibfnamefont {Snir}\ \bibnamefont {Gazit}}, \
  and\ \bibinfo {author} {\bibfnamefont {Thomas}\ \bibnamefont {Scaffidi}},\
  }\bibfield  {title} {\enquote {\bibinfo {title} {{From trivial to topological
  paramagnets: The case of ${\mathbb{Z}}_{2}$ and ${\mathbb{Z}}_{2}^{3}$
  symmetries in two dimensions}},}\ }\href {\doibase
  10.1103/PhysRevB.103.144437} {\bibfield  {journal} {\bibinfo  {journal}
  {Phys. Rev. B}\ }\textbf {\bibinfo {volume} {103}},\ \bibinfo {pages}
  {144437} (\bibinfo {year} {2021}{\natexlab{a}})}\BibitemShut {NoStop}%
\bibitem [{\citenamefont {Dupont}\ \emph
  {et~al.}(2021{\natexlab{b}})\citenamefont {Dupont}, \citenamefont {Gazit},\
  and\ \citenamefont {Scaffidi}}]{Dupont21b}%
  \BibitemOpen
  \bibfield  {author} {\bibinfo {author} {\bibfnamefont {Maxime}\ \bibnamefont
  {Dupont}}, \bibinfo {author} {\bibfnamefont {Snir}\ \bibnamefont {Gazit}}, \
  and\ \bibinfo {author} {\bibfnamefont {Thomas}\ \bibnamefont {Scaffidi}},\
  }\bibfield  {title} {\enquote {\bibinfo {title} {Evidence for deconfined
  $u(1)$ gauge theory at the transition between toric code and double
  semion},}\ }\href {\doibase 10.1103/PhysRevB.103.L140412} {\bibfield
  {journal} {\bibinfo  {journal} {Phys. Rev. B}\ }\textbf {\bibinfo {volume}
  {103}},\ \bibinfo {pages} {L140412} (\bibinfo {year}
  {2021}{\natexlab{b}})}\BibitemShut {NoStop}%
\bibitem [{\citenamefont {Tantivasadakarn}\ \emph
  {et~al.}(2023{\natexlab{a}})\citenamefont {Tantivasadakarn}, \citenamefont
  {Thorngren}, \citenamefont {Vishwanath},\ and\ \citenamefont
  {Verresen}}]{Tantivasadakarn21}%
  \BibitemOpen
  \bibfield  {author} {\bibinfo {author} {\bibfnamefont {Nathanan}\
  \bibnamefont {Tantivasadakarn}}, \bibinfo {author} {\bibfnamefont {Ryan}\
  \bibnamefont {Thorngren}}, \bibinfo {author} {\bibfnamefont {Ashvin}\
  \bibnamefont {Vishwanath}}, \ and\ \bibinfo {author} {\bibfnamefont {Ruben}\
  \bibnamefont {Verresen}},\ }\bibfield  {title} {\enquote {\bibinfo {title}
  {{Building models of topological quantum criticality from pivot
  Hamiltonians}},}\ }\href {\doibase 10.21468/SciPostPhys.14.2.013} {\bibfield
  {journal} {\bibinfo  {journal} {SciPost Phys.}\ }\textbf {\bibinfo {volume}
  {14}},\ \bibinfo {pages} {013} (\bibinfo {year}
  {2023}{\natexlab{a}})}\BibitemShut {NoStop}%
\bibitem [{\citenamefont {Tantivasadakarn}\ \emph
  {et~al.}(2023{\natexlab{b}})\citenamefont {Tantivasadakarn}, \citenamefont
  {Thorngren}, \citenamefont {Vishwanath},\ and\ \citenamefont
  {Verresen}}]{pivots}%
  \BibitemOpen
  \bibfield  {author} {\bibinfo {author} {\bibfnamefont {Nathanan}\
  \bibnamefont {Tantivasadakarn}}, \bibinfo {author} {\bibfnamefont {Ryan}\
  \bibnamefont {Thorngren}}, \bibinfo {author} {\bibfnamefont {Ashvin}\
  \bibnamefont {Vishwanath}}, \ and\ \bibinfo {author} {\bibfnamefont {Ruben}\
  \bibnamefont {Verresen}},\ }\bibfield  {title} {\enquote {\bibinfo {title}
  {{Pivot Hamiltonians as generators of symmetry and entanglement}},}\ }\href
  {\doibase 10.21468/SciPostPhys.14.2.012} {\bibfield  {journal} {\bibinfo
  {journal} {SciPost Phys.}\ }\textbf {\bibinfo {volume} {14}},\ \bibinfo
  {pages} {012} (\bibinfo {year} {2023}{\natexlab{b}})}\BibitemShut {NoStop}%
\bibitem [{\citenamefont {Suzuki}(1971)}]{Suzuki71}%
  \BibitemOpen
  \bibfield  {author} {\bibinfo {author} {\bibfnamefont {Masuo}\ \bibnamefont
  {Suzuki}},\ }\bibfield  {title} {\enquote {\bibinfo {title} {{Relationship
  among Exactly Soluble Models of Critical Phenomena. I*)2D Ising Model, Dimer
  Problem and the Generalized XY-Model}},}\ }\href {\doibase
  10.1143/PTP.46.1337} {\bibfield  {journal} {\bibinfo  {journal} {Progress of
  Theoretical Physics}\ }\textbf {\bibinfo {volume} {46}},\ \bibinfo {pages}
  {1337} (\bibinfo {year} {1971})}\BibitemShut {NoStop}%
\bibitem [{\citenamefont {Briegel}\ and\ \citenamefont
  {Raussendorf}(2001)}]{Briegel01}%
  \BibitemOpen
  \bibfield  {author} {\bibinfo {author} {\bibfnamefont {Hans~J.}\ \bibnamefont
  {Briegel}}\ and\ \bibinfo {author} {\bibfnamefont {Robert}\ \bibnamefont
  {Raussendorf}},\ }\bibfield  {title} {\enquote {\bibinfo {title} {Persistent
  entanglement in arrays of interacting particles},}\ }\href {\doibase
  10.1103/PhysRevLett.86.910} {\bibfield  {journal} {\bibinfo  {journal} {Phys.
  Rev. Lett.}\ }\textbf {\bibinfo {volume} {86}},\ \bibinfo {pages} {910--913}
  (\bibinfo {year} {2001})}\BibitemShut {NoStop}%
\bibitem [{\citenamefont {Son}\ \emph {et~al.}(2011)\citenamefont {Son},
  \citenamefont {Amico}, \citenamefont {Fazio}, \citenamefont {Hamma},
  \citenamefont {Pascazio},\ and\ \citenamefont {Vedral}}]{Son11}%
  \BibitemOpen
  \bibfield  {author} {\bibinfo {author} {\bibfnamefont {W.}~\bibnamefont
  {Son}}, \bibinfo {author} {\bibfnamefont {L.}~\bibnamefont {Amico}}, \bibinfo
  {author} {\bibfnamefont {R.}~\bibnamefont {Fazio}}, \bibinfo {author}
  {\bibfnamefont {A.}~\bibnamefont {Hamma}}, \bibinfo {author} {\bibfnamefont
  {S.}~\bibnamefont {Pascazio}}, \ and\ \bibinfo {author} {\bibfnamefont
  {V.}~\bibnamefont {Vedral}},\ }\bibfield  {title} {\enquote {\bibinfo {title}
  {Quantum phase transition between cluster and antiferromagnetic states},}\
  }\href {\doibase 10.1209/0295-5075/95/50001} {\bibfield  {journal} {\bibinfo
  {journal} {{EPL} (Europhysics Letters)}\ }\textbf {\bibinfo {volume} {95}},\
  \bibinfo {pages} {50001} (\bibinfo {year} {2011})}\BibitemShut {NoStop}%
\bibitem [{\citenamefont {Parker}\ \emph {et~al.}(2019)\citenamefont {Parker},
  \citenamefont {Vasseur},\ and\ \citenamefont {Scaffidi}}]{Parker18b}%
  \BibitemOpen
  \bibfield  {author} {\bibinfo {author} {\bibfnamefont {Daniel~E.}\
  \bibnamefont {Parker}}, \bibinfo {author} {\bibfnamefont {Romain}\
  \bibnamefont {Vasseur}}, \ and\ \bibinfo {author} {\bibfnamefont {Thomas}\
  \bibnamefont {Scaffidi}},\ }\bibfield  {title} {\enquote {\bibinfo {title}
  {{Topologically Protected Long Edge Coherence Times in Symmetry-Broken
  Phases}},}\ }\href {\doibase 10.1103/PhysRevLett.122.240605} {\bibfield
  {journal} {\bibinfo  {journal} {Phys. Rev. Lett.}\ }\textbf {\bibinfo
  {volume} {122}},\ \bibinfo {pages} {240605} (\bibinfo {year}
  {2019})}\BibitemShut {NoStop}%
\bibitem [{\citenamefont {Verresen}(2020)}]{verresen2020topology}%
  \BibitemOpen
  \bibfield  {author} {\bibinfo {author} {\bibfnamefont {Ruben}\ \bibnamefont
  {Verresen}},\ }\href@noop {} {\enquote {\bibinfo {title} {Topology and edge
  states survive quantum criticality between topological insulators},}\ }
  (\bibinfo {year} {2020}),\ \Eprint {http://arxiv.org/abs/2003.05453}
  {arXiv:2003.05453 [cond-mat.str-el]} \BibitemShut {NoStop}%
\bibitem [{\citenamefont {Geraedts}\ and\ \citenamefont
  {Motrunich}(2014)}]{geraedtsmotrunich}%
  \BibitemOpen
  \bibfield  {author} {\bibinfo {author} {\bibfnamefont {Scott~D.}\
  \bibnamefont {Geraedts}}\ and\ \bibinfo {author} {\bibfnamefont {Olexei~I.}\
  \bibnamefont {Motrunich}},\ }\href {\doibase 10.48550/ARXIV.1410.1580}
  {\enquote {\bibinfo {title} {Exact models for symmetry-protected topological
  phases in one dimension},}\ } (\bibinfo {year} {2014})\BibitemShut {NoStop}%
\bibitem [{\citenamefont {Senthil}\ \emph
  {et~al.}(2004{\natexlab{a}})\citenamefont {Senthil}, \citenamefont
  {Vishwanath}, \citenamefont {Balents}, \citenamefont {Sachdev},\ and\
  \citenamefont {Fisher}}]{Senthil04}%
  \BibitemOpen
  \bibfield  {author} {\bibinfo {author} {\bibfnamefont {T.}~\bibnamefont
  {Senthil}}, \bibinfo {author} {\bibfnamefont {Ashvin}\ \bibnamefont
  {Vishwanath}}, \bibinfo {author} {\bibfnamefont {Leon}\ \bibnamefont
  {Balents}}, \bibinfo {author} {\bibfnamefont {Subir}\ \bibnamefont
  {Sachdev}}, \ and\ \bibinfo {author} {\bibfnamefont {Matthew P.~A.}\
  \bibnamefont {Fisher}},\ }\bibfield  {title} {\enquote {\bibinfo {title}
  {{Deconfined Quantum Critical Points}},}\ }\href {\doibase
  10.1126/science.1091806} {\bibfield  {journal} {\bibinfo  {journal}
  {Science}\ }\textbf {\bibinfo {volume} {303}},\ \bibinfo {pages} {1490--1494}
  (\bibinfo {year} {2004}{\natexlab{a}})},\ \Eprint
  {http://arxiv.org/abs/https://science.sciencemag.org/content/303/5663/1490.full.pdf}
  {https://science.sciencemag.org/content/303/5663/1490.full.pdf} \BibitemShut
  {NoStop}%
\bibitem [{\citenamefont {Senthil}\ \emph
  {et~al.}(2004{\natexlab{b}})\citenamefont {Senthil}, \citenamefont {Balents},
  \citenamefont {Sachdev}, \citenamefont {Vishwanath},\ and\ \citenamefont
  {Fisher}}]{Senthil04b}%
  \BibitemOpen
  \bibfield  {author} {\bibinfo {author} {\bibfnamefont {T.}~\bibnamefont
  {Senthil}}, \bibinfo {author} {\bibfnamefont {Leon}\ \bibnamefont {Balents}},
  \bibinfo {author} {\bibfnamefont {Subir}\ \bibnamefont {Sachdev}}, \bibinfo
  {author} {\bibfnamefont {Ashvin}\ \bibnamefont {Vishwanath}}, \ and\ \bibinfo
  {author} {\bibfnamefont {Matthew P.~A.}\ \bibnamefont {Fisher}},\ }\bibfield
  {title} {\enquote {\bibinfo {title} {Quantum criticality beyond the
  landau-ginzburg-wilson paradigm},}\ }\href {\doibase
  10.1103/PhysRevB.70.144407} {\bibfield  {journal} {\bibinfo  {journal} {Phys.
  Rev. B}\ }\textbf {\bibinfo {volume} {70}},\ \bibinfo {pages} {144407}
  (\bibinfo {year} {2004}{\natexlab{b}})}\BibitemShut {NoStop}%
\bibitem [{\citenamefont {Levin}\ and\ \citenamefont
  {Senthil}(2004)}]{Levin_2004}%
  \BibitemOpen
  \bibfield  {author} {\bibinfo {author} {\bibfnamefont {Michael}\ \bibnamefont
  {Levin}}\ and\ \bibinfo {author} {\bibfnamefont {T.}~\bibnamefont
  {Senthil}},\ }\bibfield  {title} {\enquote {\bibinfo {title} {Deconfined
  quantum criticality and n{\'{e} }el order via dimer disorder},}\ }\href
  {\doibase 10.1103/physrevb.70.220403} {\bibfield  {journal} {\bibinfo
  {journal} {Physical Review B}\ }\textbf {\bibinfo {volume} {70}} (\bibinfo
  {year} {2004}),\ 10.1103/physrevb.70.220403}\BibitemShut {NoStop}%
\bibitem [{\citenamefont {Balents}\ \emph {et~al.}(2005)\citenamefont
  {Balents}, \citenamefont {Bartosch}, \citenamefont {Burkov}, \citenamefont
  {Sachdev},\ and\ \citenamefont {Sengupta}}]{Balents05}%
  \BibitemOpen
  \bibfield  {author} {\bibinfo {author} {\bibfnamefont {Leon}\ \bibnamefont
  {Balents}}, \bibinfo {author} {\bibfnamefont {Lorenz}\ \bibnamefont
  {Bartosch}}, \bibinfo {author} {\bibfnamefont {Anton}\ \bibnamefont
  {Burkov}}, \bibinfo {author} {\bibfnamefont {Subir}\ \bibnamefont {Sachdev}},
  \ and\ \bibinfo {author} {\bibfnamefont {Krishnendu}\ \bibnamefont
  {Sengupta}},\ }\bibfield  {title} {\enquote {\bibinfo {title} {Putting
  competing orders in their place near the mott transition},}\ }\href {\doibase
  10.1103/PhysRevB.71.144508} {\bibfield  {journal} {\bibinfo  {journal} {Phys.
  Rev. B}\ }\textbf {\bibinfo {volume} {71}},\ \bibinfo {pages} {144508}
  (\bibinfo {year} {2005})}\BibitemShut {NoStop}%
\bibitem [{\citenamefont {Vishwanath}\ \emph {et~al.}(2004)\citenamefont
  {Vishwanath}, \citenamefont {Balents},\ and\ \citenamefont
  {Senthil}}]{Vishwanath04}%
  \BibitemOpen
  \bibfield  {author} {\bibinfo {author} {\bibfnamefont {Ashvin}\ \bibnamefont
  {Vishwanath}}, \bibinfo {author} {\bibfnamefont {L.}~\bibnamefont {Balents}},
  \ and\ \bibinfo {author} {\bibfnamefont {T.}~\bibnamefont {Senthil}},\
  }\bibfield  {title} {\enquote {\bibinfo {title} {Quantum criticality and
  deconfinement in phase transitions between valence bond solids},}\ }\href
  {\doibase 10.1103/PhysRevB.69.224416} {\bibfield  {journal} {\bibinfo
  {journal} {Phys. Rev. B}\ }\textbf {\bibinfo {volume} {69}},\ \bibinfo
  {pages} {224416} (\bibinfo {year} {2004})}\BibitemShut {NoStop}%
\bibitem [{\citenamefont {Ghaemi}\ and\ \citenamefont
  {Senthil}(2006)}]{Ghaemi06}%
  \BibitemOpen
  \bibfield  {author} {\bibinfo {author} {\bibfnamefont {Pouyan}\ \bibnamefont
  {Ghaemi}}\ and\ \bibinfo {author} {\bibfnamefont {T.}~\bibnamefont
  {Senthil}},\ }\bibfield  {title} {\enquote {\bibinfo {title} {N\'eel order,
  quantum spin liquids, and quantum criticality in two dimensions},}\ }\href
  {\doibase 10.1103/PhysRevB.73.054415} {\bibfield  {journal} {\bibinfo
  {journal} {Phys. Rev. B}\ }\textbf {\bibinfo {volume} {73}},\ \bibinfo
  {pages} {054415} (\bibinfo {year} {2006})}\BibitemShut {NoStop}%
\bibitem [{\citenamefont {Sandvik}(2007)}]{Sandvik07b}%
  \BibitemOpen
  \bibfield  {author} {\bibinfo {author} {\bibfnamefont {Anders~W.}\
  \bibnamefont {Sandvik}},\ }\bibfield  {title} {\enquote {\bibinfo {title}
  {Evidence for deconfined quantum criticality in a two-dimensional heisenberg
  model with four-spin interactions},}\ }\href {\doibase
  10.1103/PhysRevLett.98.227202} {\bibfield  {journal} {\bibinfo  {journal}
  {Phys. Rev. Lett.}\ }\textbf {\bibinfo {volume} {98}},\ \bibinfo {pages}
  {227202} (\bibinfo {year} {2007})}\BibitemShut {NoStop}%
\bibitem [{\citenamefont {Grover}\ and\ \citenamefont
  {Senthil}(2007)}]{Grover07}%
  \BibitemOpen
  \bibfield  {author} {\bibinfo {author} {\bibfnamefont {Tarun}\ \bibnamefont
  {Grover}}\ and\ \bibinfo {author} {\bibfnamefont {T.}~\bibnamefont
  {Senthil}},\ }\bibfield  {title} {\enquote {\bibinfo {title} {Quantum spin
  nematics, dimerization, and deconfined criticality in quasi-1d spin-one
  magnets},}\ }\href {\doibase 10.1103/PhysRevLett.98.247202} {\bibfield
  {journal} {\bibinfo  {journal} {Phys. Rev. Lett.}\ }\textbf {\bibinfo
  {volume} {98}},\ \bibinfo {pages} {247202} (\bibinfo {year}
  {2007})}\BibitemShut {NoStop}%
\bibitem [{\citenamefont {Melko}\ and\ \citenamefont {Kaul}(2008)}]{Melko08}%
  \BibitemOpen
  \bibfield  {author} {\bibinfo {author} {\bibfnamefont {Roger~G.}\
  \bibnamefont {Melko}}\ and\ \bibinfo {author} {\bibfnamefont {Ribhu~K.}\
  \bibnamefont {Kaul}},\ }\bibfield  {title} {\enquote {\bibinfo {title}
  {Scaling in the fan of an unconventional quantum critical point},}\ }\href
  {\doibase 10.1103/PhysRevLett.100.017203} {\bibfield  {journal} {\bibinfo
  {journal} {Phys. Rev. Lett.}\ }\textbf {\bibinfo {volume} {100}},\ \bibinfo
  {pages} {017203} (\bibinfo {year} {2008})}\BibitemShut {NoStop}%
\bibitem [{\citenamefont {Sandvik}(2010)}]{Sandvik10}%
  \BibitemOpen
  \bibfield  {author} {\bibinfo {author} {\bibfnamefont {Anders~W.}\
  \bibnamefont {Sandvik}},\ }\bibfield  {title} {\enquote {\bibinfo {title}
  {Continuous quantum phase transition between an antiferromagnet and a
  valence-bond solid in two dimensions: Evidence for logarithmic corrections to
  scaling},}\ }\href {\doibase 10.1103/PhysRevLett.104.177201} {\bibfield
  {journal} {\bibinfo  {journal} {Phys. Rev. Lett.}\ }\textbf {\bibinfo
  {volume} {104}},\ \bibinfo {pages} {177201} (\bibinfo {year}
  {2010})}\BibitemShut {NoStop}%
\bibitem [{\citenamefont {Chen}\ \emph
  {et~al.}(2013{\natexlab{b}})\citenamefont {Chen}, \citenamefont {Huang},
  \citenamefont {Deng}, \citenamefont {Kuklov}, \citenamefont {Prokof'ev},\
  and\ \citenamefont {Svistunov}}]{KunChen13}%
  \BibitemOpen
  \bibfield  {author} {\bibinfo {author} {\bibfnamefont {Kun}\ \bibnamefont
  {Chen}}, \bibinfo {author} {\bibfnamefont {Yuan}\ \bibnamefont {Huang}},
  \bibinfo {author} {\bibfnamefont {Youjin}\ \bibnamefont {Deng}}, \bibinfo
  {author} {\bibfnamefont {A.~B.}\ \bibnamefont {Kuklov}}, \bibinfo {author}
  {\bibfnamefont {N.~V.}\ \bibnamefont {Prokof'ev}}, \ and\ \bibinfo {author}
  {\bibfnamefont {B.~V.}\ \bibnamefont {Svistunov}},\ }\bibfield  {title}
  {\enquote {\bibinfo {title} {Deconfined criticality flow in the heisenberg
  model with ring-exchange interactions},}\ }\href {\doibase
  10.1103/PhysRevLett.110.185701} {\bibfield  {journal} {\bibinfo  {journal}
  {Phys. Rev. Lett.}\ }\textbf {\bibinfo {volume} {110}},\ \bibinfo {pages}
  {185701} (\bibinfo {year} {2013}{\natexlab{b}})}\BibitemShut {NoStop}%
\bibitem [{\citenamefont {Nahum}\ \emph {et~al.}(2015)\citenamefont {Nahum},
  \citenamefont {Serna}, \citenamefont {Chalker}, \citenamefont {Ortu\~no},\
  and\ \citenamefont {Somoza}}]{Nahum15}%
  \BibitemOpen
  \bibfield  {author} {\bibinfo {author} {\bibfnamefont {Adam}\ \bibnamefont
  {Nahum}}, \bibinfo {author} {\bibfnamefont {P.}~\bibnamefont {Serna}},
  \bibinfo {author} {\bibfnamefont {J.~T.}\ \bibnamefont {Chalker}}, \bibinfo
  {author} {\bibfnamefont {M.}~\bibnamefont {Ortu\~no}}, \ and\ \bibinfo
  {author} {\bibfnamefont {A.~M.}\ \bibnamefont {Somoza}},\ }\bibfield  {title}
  {\enquote {\bibinfo {title} {Emergent so(5) symmetry at the n\'eel to
  valence-bond-solid transition},}\ }\href {\doibase
  10.1103/PhysRevLett.115.267203} {\bibfield  {journal} {\bibinfo  {journal}
  {Phys. Rev. Lett.}\ }\textbf {\bibinfo {volume} {115}},\ \bibinfo {pages}
  {267203} (\bibinfo {year} {2015})}\BibitemShut {NoStop}%
\bibitem [{\citenamefont {Shao}\ \emph {et~al.}(2016)\citenamefont {Shao},
  \citenamefont {Guo},\ and\ \citenamefont {Sandvik}}]{Shao16}%
  \BibitemOpen
  \bibfield  {author} {\bibinfo {author} {\bibfnamefont {Hui}\ \bibnamefont
  {Shao}}, \bibinfo {author} {\bibfnamefont {Wenan}\ \bibnamefont {Guo}}, \
  and\ \bibinfo {author} {\bibfnamefont {Anders~W.}\ \bibnamefont {Sandvik}},\
  }\bibfield  {title} {\enquote {\bibinfo {title} {Quantum criticality with two
  length scales},}\ }\href {\doibase 10.1126/science.aad5007} {\bibfield
  {journal} {\bibinfo  {journal} {Science}\ }\textbf {\bibinfo {volume}
  {352}},\ \bibinfo {pages} {213--216} (\bibinfo {year} {2016})}\BibitemShut
  {NoStop}%
\bibitem [{\citenamefont {Wang}\ \emph
  {et~al.}(2017{\natexlab{b}})\citenamefont {Wang}, \citenamefont {Nahum},
  \citenamefont {Metlitski}, \citenamefont {Xu},\ and\ \citenamefont
  {Senthil}}]{Wang_2017}%
  \BibitemOpen
  \bibfield  {author} {\bibinfo {author} {\bibfnamefont {Chong}\ \bibnamefont
  {Wang}}, \bibinfo {author} {\bibfnamefont {Adam}\ \bibnamefont {Nahum}},
  \bibinfo {author} {\bibfnamefont {Max~A.}\ \bibnamefont {Metlitski}},
  \bibinfo {author} {\bibfnamefont {Cenke}\ \bibnamefont {Xu}}, \ and\ \bibinfo
  {author} {\bibfnamefont {T.}~\bibnamefont {Senthil}},\ }\bibfield  {title}
  {\enquote {\bibinfo {title} {Deconfined quantum critical points: Symmetries
  and dualities},}\ }\href {\doibase 10.1103/physrevx.7.031051} {\bibfield
  {journal} {\bibinfo  {journal} {Physical Review X}\ }\textbf {\bibinfo
  {volume} {7}} (\bibinfo {year} {2017}{\natexlab{b}}),\
  10.1103/physrevx.7.031051}\BibitemShut {NoStop}%
\bibitem [{\citenamefont {Ma}\ \emph {et~al.}(2018)\citenamefont {Ma},
  \citenamefont {Sun}, \citenamefont {You}, \citenamefont {Xu}, \citenamefont
  {Vishwanath}, \citenamefont {Sandvik},\ and\ \citenamefont {Meng}}]{Ma18}%
  \BibitemOpen
  \bibfield  {author} {\bibinfo {author} {\bibfnamefont {Nvsen}\ \bibnamefont
  {Ma}}, \bibinfo {author} {\bibfnamefont {Guang-Yu}\ \bibnamefont {Sun}},
  \bibinfo {author} {\bibfnamefont {Yi-Zhuang}\ \bibnamefont {You}}, \bibinfo
  {author} {\bibfnamefont {Cenke}\ \bibnamefont {Xu}}, \bibinfo {author}
  {\bibfnamefont {Ashvin}\ \bibnamefont {Vishwanath}}, \bibinfo {author}
  {\bibfnamefont {Anders~W.}\ \bibnamefont {Sandvik}}, \ and\ \bibinfo {author}
  {\bibfnamefont {Zi~Yang}\ \bibnamefont {Meng}},\ }\bibfield  {title}
  {\enquote {\bibinfo {title} {Dynamical signature of fractionalization at a
  deconfined quantum critical point},}\ }\href {\doibase
  10.1103/PhysRevB.98.174421} {\bibfield  {journal} {\bibinfo  {journal} {Phys.
  Rev. B}\ }\textbf {\bibinfo {volume} {98}},\ \bibinfo {pages} {174421}
  (\bibinfo {year} {2018})}\BibitemShut {NoStop}%
\bibitem [{\citenamefont {Ma}\ \emph {et~al.}(2019)\citenamefont {Ma},
  \citenamefont {You},\ and\ \citenamefont {Meng}}]{Ma19}%
  \BibitemOpen
  \bibfield  {author} {\bibinfo {author} {\bibfnamefont {Nvsen}\ \bibnamefont
  {Ma}}, \bibinfo {author} {\bibfnamefont {Yi-Zhuang}\ \bibnamefont {You}}, \
  and\ \bibinfo {author} {\bibfnamefont {Zi~Yang}\ \bibnamefont {Meng}},\
  }\bibfield  {title} {\enquote {\bibinfo {title} {Role of noether's theorem at
  the deconfined quantum critical point},}\ }\href {\doibase
  10.1103/PhysRevLett.122.175701} {\bibfield  {journal} {\bibinfo  {journal}
  {Phys. Rev. Lett.}\ }\textbf {\bibinfo {volume} {122}},\ \bibinfo {pages}
  {175701} (\bibinfo {year} {2019})}\BibitemShut {NoStop}%
\bibitem [{\citenamefont {Sreejith}\ \emph {et~al.}(2019)\citenamefont
  {Sreejith}, \citenamefont {Powell},\ and\ \citenamefont
  {Nahum}}]{Sreejith19}%
  \BibitemOpen
  \bibfield  {author} {\bibinfo {author} {\bibfnamefont {G.~J.}\ \bibnamefont
  {Sreejith}}, \bibinfo {author} {\bibfnamefont {Stephen}\ \bibnamefont
  {Powell}}, \ and\ \bibinfo {author} {\bibfnamefont {Adam}\ \bibnamefont
  {Nahum}},\ }\bibfield  {title} {\enquote {\bibinfo {title} {Emergent so(5)
  symmetry at the columnar ordering transition in the classical cubic dimer
  model},}\ }\href {\doibase 10.1103/PhysRevLett.122.080601} {\bibfield
  {journal} {\bibinfo  {journal} {Phys. Rev. Lett.}\ }\textbf {\bibinfo
  {volume} {122}},\ \bibinfo {pages} {080601} (\bibinfo {year}
  {2019})}\BibitemShut {NoStop}%
\bibitem [{\citenamefont {Li}\ \emph {et~al.}(2019)\citenamefont {Li},
  \citenamefont {Jian},\ and\ \citenamefont {Yao}}]{Li19}%
  \BibitemOpen
  \bibfield  {author} {\bibinfo {author} {\bibfnamefont {Zi-Xiang}\
  \bibnamefont {Li}}, \bibinfo {author} {\bibfnamefont {Shao-Kai}\ \bibnamefont
  {Jian}}, \ and\ \bibinfo {author} {\bibfnamefont {Hong}\ \bibnamefont
  {Yao}},\ }\href {\doibase 10.48550/ARXIV.1904.10975} {\enquote {\bibinfo
  {title} {Deconfined quantum criticality and emergent so(5) symmetry in
  fermionic systems},}\ } (\bibinfo {year} {2019})\BibitemShut {NoStop}%
\bibitem [{\citenamefont {Takahashi}\ and\ \citenamefont
  {Sandvik}(2020)}]{Takahashi20}%
  \BibitemOpen
  \bibfield  {author} {\bibinfo {author} {\bibfnamefont {Jun}\ \bibnamefont
  {Takahashi}}\ and\ \bibinfo {author} {\bibfnamefont {Anders~W.}\ \bibnamefont
  {Sandvik}},\ }\bibfield  {title} {\enquote {\bibinfo {title} {Valence-bond
  solids, vestigial order, and emergent so(5) symmetry in a two-dimensional
  quantum magnet},}\ }\href {\doibase 10.1103/PhysRevResearch.2.033459}
  {\bibfield  {journal} {\bibinfo  {journal} {Phys. Rev. Research}\ }\textbf
  {\bibinfo {volume} {2}},\ \bibinfo {pages} {033459} (\bibinfo {year}
  {2020})}\BibitemShut {NoStop}%
\bibitem [{\citenamefont {Wang}\ \emph {et~al.}(2021)\citenamefont {Wang},
  \citenamefont {Zaletel}, \citenamefont {Mong},\ and\ \citenamefont
  {Assaad}}]{Wang21}%
  \BibitemOpen
  \bibfield  {author} {\bibinfo {author} {\bibfnamefont {Zhenjiu}\ \bibnamefont
  {Wang}}, \bibinfo {author} {\bibfnamefont {Michael~P.}\ \bibnamefont
  {Zaletel}}, \bibinfo {author} {\bibfnamefont {Roger S.~K.}\ \bibnamefont
  {Mong}}, \ and\ \bibinfo {author} {\bibfnamefont {Fakher~F.}\ \bibnamefont
  {Assaad}},\ }\bibfield  {title} {\enquote {\bibinfo {title} {Phases of the
  ($2+1$) dimensional so(5) nonlinear sigma model with topological term},}\
  }\href {\doibase 10.1103/PhysRevLett.126.045701} {\bibfield  {journal}
  {\bibinfo  {journal} {Phys. Rev. Lett.}\ }\textbf {\bibinfo {volume} {126}},\
  \bibinfo {pages} {045701} (\bibinfo {year} {2021})}\BibitemShut {NoStop}%
\bibitem [{\citenamefont {Ogino}\ \emph {et~al.}(2021)\citenamefont {Ogino},
  \citenamefont {Kaneko}, \citenamefont {Morita}, \citenamefont {Furukawa},\
  and\ \citenamefont {Kawashima}}]{Ogino21}%
  \BibitemOpen
  \bibfield  {author} {\bibinfo {author} {\bibfnamefont {Takuhiro}\
  \bibnamefont {Ogino}}, \bibinfo {author} {\bibfnamefont {Ryui}\ \bibnamefont
  {Kaneko}}, \bibinfo {author} {\bibfnamefont {Satoshi}\ \bibnamefont
  {Morita}}, \bibinfo {author} {\bibfnamefont {Shunsuke}\ \bibnamefont
  {Furukawa}}, \ and\ \bibinfo {author} {\bibfnamefont {Naoki}\ \bibnamefont
  {Kawashima}},\ }\bibfield  {title} {\enquote {\bibinfo {title} {Continuous
  phase transition between n\'eel and valence bond solid phases in a
  $j\text{\ensuremath{-}}q$-like spin ladder system},}\ }\href {\doibase
  10.1103/PhysRevB.103.085117} {\bibfield  {journal} {\bibinfo  {journal}
  {Phys. Rev. B}\ }\textbf {\bibinfo {volume} {103}},\ \bibinfo {pages}
  {085117} (\bibinfo {year} {2021})}\BibitemShut {NoStop}%
\bibitem [{\citenamefont {Roberts}\ \emph {et~al.}(2019)\citenamefont
  {Roberts}, \citenamefont {Jiang},\ and\ \citenamefont
  {Motrunich}}]{Roberts19}%
  \BibitemOpen
  \bibfield  {author} {\bibinfo {author} {\bibfnamefont {Brenden}\ \bibnamefont
  {Roberts}}, \bibinfo {author} {\bibfnamefont {Shenghan}\ \bibnamefont
  {Jiang}}, \ and\ \bibinfo {author} {\bibfnamefont {Olexei~I.}\ \bibnamefont
  {Motrunich}},\ }\bibfield  {title} {\enquote {\bibinfo {title} {Deconfined
  quantum critical point in one dimension},}\ }\href {\doibase
  10.1103/PhysRevB.99.165143} {\bibfield  {journal} {\bibinfo  {journal} {Phys.
  Rev. B}\ }\textbf {\bibinfo {volume} {99}},\ \bibinfo {pages} {165143}
  (\bibinfo {year} {2019})}\BibitemShut {NoStop}%
\bibitem [{\citenamefont {Huang}\ \emph {et~al.}(2019)\citenamefont {Huang},
  \citenamefont {Lu}, \citenamefont {You}, \citenamefont {Meng},\ and\
  \citenamefont {Xiang}}]{Huang_2019}%
  \BibitemOpen
  \bibfield  {author} {\bibinfo {author} {\bibfnamefont {Rui-Zhen}\
  \bibnamefont {Huang}}, \bibinfo {author} {\bibfnamefont {Da-Chuan}\
  \bibnamefont {Lu}}, \bibinfo {author} {\bibfnamefont {Yi-Zhuang}\
  \bibnamefont {You}}, \bibinfo {author} {\bibfnamefont {Zi~Yang}\ \bibnamefont
  {Meng}}, \ and\ \bibinfo {author} {\bibfnamefont {Tao}\ \bibnamefont
  {Xiang}},\ }\bibfield  {title} {\enquote {\bibinfo {title} {Emergent symmetry
  and conserved current at a one-dimensional incarnation of deconfined quantum
  critical point},}\ }\href {\doibase 10.1103/physrevb.100.125137} {\bibfield
  {journal} {\bibinfo  {journal} {Physical Review B}\ }\textbf {\bibinfo
  {volume} {100}} (\bibinfo {year} {2019}),\
  10.1103/physrevb.100.125137}\BibitemShut {NoStop}%
\bibitem [{\citenamefont {Mudry}\ \emph {et~al.}(2019)\citenamefont {Mudry},
  \citenamefont {Furusaki}, \citenamefont {Morimoto},\ and\ \citenamefont
  {Hikihara}}]{Mudry_2019}%
  \BibitemOpen
  \bibfield  {author} {\bibinfo {author} {\bibfnamefont {Christopher}\
  \bibnamefont {Mudry}}, \bibinfo {author} {\bibfnamefont {Akira}\ \bibnamefont
  {Furusaki}}, \bibinfo {author} {\bibfnamefont {Takahiro}\ \bibnamefont
  {Morimoto}}, \ and\ \bibinfo {author} {\bibfnamefont {Toshiya}\ \bibnamefont
  {Hikihara}},\ }\bibfield  {title} {\enquote {\bibinfo {title} {Quantum phase
  transitions beyond landau-ginzburg theory in one-dimensional space
  revisited},}\ }\href {\doibase 10.1103/physrevb.99.205153} {\bibfield
  {journal} {\bibinfo  {journal} {Physical Review B}\ }\textbf {\bibinfo
  {volume} {99}} (\bibinfo {year} {2019}),\
  10.1103/physrevb.99.205153}\BibitemShut {NoStop}%
\bibitem [{\citenamefont {Jiang}\ and\ \citenamefont
  {Motrunich}(2019)}]{Jiang19}%
  \BibitemOpen
  \bibfield  {author} {\bibinfo {author} {\bibfnamefont {Shenghan}\
  \bibnamefont {Jiang}}\ and\ \bibinfo {author} {\bibfnamefont {Olexei}\
  \bibnamefont {Motrunich}},\ }\bibfield  {title} {\enquote {\bibinfo {title}
  {{Ising ferromagnet to valence bond solid transition in a one-dimensional
  spin chain: Analogies to deconfined quantum critical points}},}\ }\href
  {\doibase 10.1103/PhysRevB.99.075103} {\bibfield  {journal} {\bibinfo
  {journal} {Phys. Rev. B}\ }\textbf {\bibinfo {volume} {99}},\ \bibinfo
  {pages} {075103} (\bibinfo {year} {2019})}\BibitemShut {NoStop}%
\bibitem [{\citenamefont {Sun}\ \emph {et~al.}(2019)\citenamefont {Sun},
  \citenamefont {Wei},\ and\ \citenamefont {Kou}}]{Sun19}%
  \BibitemOpen
  \bibfield  {author} {\bibinfo {author} {\bibfnamefont {Gaoyong}\ \bibnamefont
  {Sun}}, \bibinfo {author} {\bibfnamefont {Bo-Bo}\ \bibnamefont {Wei}}, \ and\
  \bibinfo {author} {\bibfnamefont {Su-Peng}\ \bibnamefont {Kou}},\ }\bibfield
  {title} {\enquote {\bibinfo {title} {Fidelity as a probe for a deconfined
  quantum critical point},}\ }\href {\doibase 10.1103/PhysRevB.100.064427}
  {\bibfield  {journal} {\bibinfo  {journal} {Phys. Rev. B}\ }\textbf {\bibinfo
  {volume} {100}},\ \bibinfo {pages} {064427} (\bibinfo {year}
  {2019})}\BibitemShut {NoStop}%
\bibitem [{\citenamefont {Yang}\ \emph {et~al.}(2020)\citenamefont {Yang},
  \citenamefont {Yao},\ and\ \citenamefont {Sandvik}}]{Yang20}%
  \BibitemOpen
  \bibfield  {author} {\bibinfo {author} {\bibfnamefont {Sibin}\ \bibnamefont
  {Yang}}, \bibinfo {author} {\bibfnamefont {Dao-Xin}\ \bibnamefont {Yao}}, \
  and\ \bibinfo {author} {\bibfnamefont {Anders~W.}\ \bibnamefont {Sandvik}},\
  }\href {\doibase 10.48550/ARXIV.2001.02821} {\enquote {\bibinfo {title}
  {Deconfined quantum criticality in spin-1/2 chains with long-range
  interactions},}\ } (\bibinfo {year} {2020})\BibitemShut {NoStop}%
\bibitem [{\citenamefont {Roberts}\ \emph {et~al.}(2021)\citenamefont
  {Roberts}, \citenamefont {Jiang},\ and\ \citenamefont {Motrunich}}]{RJM21}%
  \BibitemOpen
  \bibfield  {author} {\bibinfo {author} {\bibfnamefont {Brenden}\ \bibnamefont
  {Roberts}}, \bibinfo {author} {\bibfnamefont {Shenghan}\ \bibnamefont
  {Jiang}}, \ and\ \bibinfo {author} {\bibfnamefont {Olexei~I.}\ \bibnamefont
  {Motrunich}},\ }\bibfield  {title} {\enquote {\bibinfo {title}
  {One-dimensional model for deconfined criticality with $\mathbb{Z}_3\times
  \mathbb{Z}_3$ symmetry},}\ }\href {\doibase 10.1103/PhysRevB.103.155143}
  {\bibfield  {journal} {\bibinfo  {journal} {Phys. Rev. B}\ }\textbf {\bibinfo
  {volume} {103}},\ \bibinfo {pages} {155143} (\bibinfo {year}
  {2021})}\BibitemShut {NoStop}%
\bibitem [{\citenamefont {Zhang}\ and\ \citenamefont
  {Levin}(2023)}]{ZhangLevin2022}%
  \BibitemOpen
  \bibfield  {author} {\bibinfo {author} {\bibfnamefont {Carolyn}\ \bibnamefont
  {Zhang}}\ and\ \bibinfo {author} {\bibfnamefont {Michael}\ \bibnamefont
  {Levin}},\ }\bibfield  {title} {\enquote {\bibinfo {title} {Exactly solvable
  model for a deconfined quantum critical point in 1d},}\ }\href {\doibase
  10.1103/PhysRevLett.130.026801} {\bibfield  {journal} {\bibinfo  {journal}
  {Phys. Rev. Lett.}\ }\textbf {\bibinfo {volume} {130}},\ \bibinfo {pages}
  {026801} (\bibinfo {year} {2023})}\BibitemShut {NoStop}%
\bibitem [{\citenamefont {Pollmann}\ and\ \citenamefont
  {Turner}(2012)}]{Pollmann12b}%
  \BibitemOpen
  \bibfield  {author} {\bibinfo {author} {\bibfnamefont {Frank}\ \bibnamefont
  {Pollmann}}\ and\ \bibinfo {author} {\bibfnamefont {Ari~M.}\ \bibnamefont
  {Turner}},\ }\bibfield  {title} {\enquote {\bibinfo {title} {Detection of
  symmetry-protected topological phases in one dimension},}\ }\href {\doibase
  10.1103/PhysRevB.86.125441} {\bibfield  {journal} {\bibinfo  {journal} {Phys.
  Rev. B}\ }\textbf {\bibinfo {volume} {86}},\ \bibinfo {pages} {125441}
  (\bibinfo {year} {2012})}\BibitemShut {NoStop}%
\bibitem [{\citenamefont {White}(1992)}]{White92}%
  \BibitemOpen
  \bibfield  {author} {\bibinfo {author} {\bibfnamefont {Steven~R.}\
  \bibnamefont {White}},\ }\bibfield  {title} {\enquote {\bibinfo {title}
  {Density matrix formulation for quantum renormalization groups},}\ }\href
  {\doibase 10.1103/PhysRevLett.69.2863} {\bibfield  {journal} {\bibinfo
  {journal} {Phys. Rev. Lett.}\ }\textbf {\bibinfo {volume} {69}},\ \bibinfo
  {pages} {2863--2866} (\bibinfo {year} {1992})}\BibitemShut {NoStop}%
\bibitem [{\citenamefont {Hauschild}\ and\ \citenamefont
  {Pollmann}(2018)}]{Hauschild18}%
  \BibitemOpen
  \bibfield  {author} {\bibinfo {author} {\bibfnamefont {Johannes}\
  \bibnamefont {Hauschild}}\ and\ \bibinfo {author} {\bibfnamefont {Frank}\
  \bibnamefont {Pollmann}},\ }\bibfield  {title} {\enquote {\bibinfo {title}
  {{Efficient numerical simulations with Tensor Networks: Tensor Network Python
  (TeNPy)}},}\ }\href {\doibase 10.21468/SciPostPhysLectNotes.5} {\bibfield
  {journal} {\bibinfo  {journal} {SciPost Phys. Lect. Notes}\ ,\ \bibinfo
  {pages} {5}} (\bibinfo {year} {2018})}\BibitemShut {NoStop}%
\bibitem [{sup()}]{suppl}%
  \BibitemOpen
  \href@noop {} {}\bibinfo {note} {See Supplemental Material.}\BibitemShut
  {Stop}%
\bibitem [{\citenamefont {Santos}(2015)}]{Santos15}%
  \BibitemOpen
  \bibfield  {author} {\bibinfo {author} {\bibfnamefont {Luiz~H.}\ \bibnamefont
  {Santos}},\ }\bibfield  {title} {\enquote {\bibinfo {title}
  {{Rokhsar-Kivelson models of bosonic symmetry-protected topological
  states}},}\ }\href {\doibase 10.1103/PhysRevB.91.155150} {\bibfield
  {journal} {\bibinfo  {journal} {Phys. Rev. B}\ }\textbf {\bibinfo {volume}
  {91}},\ \bibinfo {pages} {155150} (\bibinfo {year} {2015})}\BibitemShut
  {NoStop}%
\bibitem [{\citenamefont {Roberts}()}]{Roberts_conversation}%
  \BibitemOpen
  \bibfield  {author} {\bibinfo {author} {\bibfnamefont {Brendan}\ \bibnamefont
  {Roberts}},\ }\href@noop {} {}\bibinfo {howpublished} {personal
  communication}\BibitemShut {NoStop}%
\bibitem [{\citenamefont {Whitsitt}\ \emph {et~al.}(2018)\citenamefont
  {Whitsitt}, \citenamefont {Samajdar},\ and\ \citenamefont
  {Sachdev}}]{Whitsitt_2018}%
  \BibitemOpen
  \bibfield  {author} {\bibinfo {author} {\bibfnamefont {Seth}\ \bibnamefont
  {Whitsitt}}, \bibinfo {author} {\bibfnamefont {Rhine}\ \bibnamefont
  {Samajdar}}, \ and\ \bibinfo {author} {\bibfnamefont {Subir}\ \bibnamefont
  {Sachdev}},\ }\bibfield  {title} {\enquote {\bibinfo {title} {Quantum field
  theory for the chiral clock transition in one spatial dimension},}\ }\href
  {\doibase 10.1103/physrevb.98.205118} {\bibfield  {journal} {\bibinfo
  {journal} {Physical Review B}\ }\textbf {\bibinfo {volume} {98}} (\bibinfo
  {year} {2018}),\ 10.1103/physrevb.98.205118}\BibitemShut {NoStop}%
\bibitem [{\citenamefont {Kitaev}(2001)}]{kitaev_unpaired_2001}%
  \BibitemOpen
  \bibfield  {author} {\bibinfo {author} {\bibfnamefont {Alexei}\ \bibnamefont
  {Kitaev}},\ }\bibfield  {title} {\enquote {\bibinfo {title} {Unpaired
  {Majorana} fermions in quantum wires},}\ }\href {\doibase
  10.1070/1063-7869/44/10S/S29} {\bibfield  {journal} {\bibinfo  {journal}
  {Physics-Uspekhi}\ }\textbf {\bibinfo {volume} {44}},\ \bibinfo {pages}
  {131--136} (\bibinfo {year} {2001})},\ \bibinfo {note} {arXiv:
  cond-mat/0010440}\BibitemShut {NoStop}%
\bibitem [{\citenamefont {Cardy}(2005)}]{cardyBCFT}%
  \BibitemOpen
  \bibfield  {author} {\bibinfo {author} {\bibfnamefont {J.}~\bibnamefont
  {Cardy}},\ }\bibfield  {title} {\enquote {\bibinfo {title} {Boundary
  conformal field theory},}\ }in\ \href@noop {} {\emph {\bibinfo {booktitle}
  {\textit{Encyclopedia of Mathematical Physics}}}},\ \bibinfo {editor} {edited
  by\ \bibinfo {editor} {\bibfnamefont {J.-P.}\ \bibnamefont {Francoise}},
  \bibinfo {editor} {\bibfnamefont {G.}~\bibnamefont {Naber}}, \ and\ \bibinfo
  {editor} {\bibfnamefont {T.~S.}\ \bibnamefont {Tsun}}}\ (\bibinfo
  {publisher} {Elsevier},\ \bibinfo {year} {2005})\BibitemShut {NoStop}%
\bibitem [{\citenamefont {Petkova}\ and\ \citenamefont
  {Zuber}(2001)}]{PetkovaZuber}%
  \BibitemOpen
  \bibfield  {author} {\bibinfo {author} {\bibfnamefont {V.B.}\ \bibnamefont
  {Petkova}}\ and\ \bibinfo {author} {\bibfnamefont {J.-B.}\ \bibnamefont
  {Zuber}},\ }\bibfield  {title} {\enquote {\bibinfo {title} {Generalised
  twisted partition functions},}\ }\href {\doibase
  10.1016/s0370-2693(01)00276-3} {\bibfield  {journal} {\bibinfo  {journal}
  {Physics Letters B}\ }\textbf {\bibinfo {volume} {504}},\ \bibinfo {pages}
  {157--164} (\bibinfo {year} {2001})}\BibitemShut {NoStop}%
\bibitem [{\citenamefont {Ginsparg}(1988)}]{Ginsparg88}%
  \BibitemOpen
  \bibfield  {author} {\bibinfo {author} {\bibfnamefont {P.}~\bibnamefont
  {Ginsparg}},\ }\bibfield  {title} {\enquote {\bibinfo {title} {Applied
  conformal field theory},}\ }in\ \href@noop {} {\emph {\bibinfo {booktitle}
  {Fields, Strings and Critical Phenomena}}},\ \bibinfo {series and number}
  {Les Houches Summer School},\ \bibinfo {editor} {edited by\ \bibinfo {editor}
  {\bibfnamefont {E.}~\bibnamefont {Brézin}}\ and\ \bibinfo {editor}
  {\bibfnamefont {J.}~\bibnamefont {Zinn-Justin}}}\ (\bibinfo  {publisher}
  {North-Holland},\ \bibinfo {address} {Amsterdam},\ \bibinfo {year}
  {1988})\BibitemShut {NoStop}%
\bibitem [{\citenamefont {Francesco}\ \emph {et~al.}(1997)\citenamefont
  {Francesco}, \citenamefont {Mathieu},\ and\ \citenamefont
  {Sénéchal}}]{francesco_conformal_1997}%
  \BibitemOpen
  \bibfield  {author} {\bibinfo {author} {\bibfnamefont {P.~Di}\ \bibnamefont
  {Francesco}}, \bibinfo {author} {\bibfnamefont {P.}~\bibnamefont {Mathieu}},
  \ and\ \bibinfo {author} {\bibfnamefont {D.}~\bibnamefont {Sénéchal}},\
  }\href@noop {} {\emph {\bibinfo {title} {\textit{Conformal Field Theory}}}},\
  Graduate Texts in Contemporary Physics\ (\bibinfo  {publisher}
  {Springer-Verlag},\ \bibinfo {address} {New York},\ \bibinfo {year}
  {1997})\BibitemShut {NoStop}%
\bibitem [{\citenamefont {Kormos}\ \emph {et~al.}(2009)\citenamefont {Kormos},
  \citenamefont {Runkel},\ and\ \citenamefont {Watts}}]{Kormos_2009}%
  \BibitemOpen
  \bibfield  {author} {\bibinfo {author} {\bibfnamefont {M{\'{a}}rton}\
  \bibnamefont {Kormos}}, \bibinfo {author} {\bibfnamefont {Ingo}\ \bibnamefont
  {Runkel}}, \ and\ \bibinfo {author} {\bibfnamefont {G{\'{e}}rard~M.T}\
  \bibnamefont {Watts}},\ }\bibfield  {title} {\enquote {\bibinfo {title}
  {Defect flows in minimal models},}\ }\href {\doibase
  10.1088/1126-6708/2009/11/057} {\bibfield  {journal} {\bibinfo  {journal}
  {Journal of High Energy Physics}\ }\textbf {\bibinfo {volume} {2009}},\
  \bibinfo {pages} {057--057} (\bibinfo {year} {2009})}\BibitemShut {NoStop}%
\bibitem [{\citenamefont {Affleck}\ \emph {et~al.}(1998)\citenamefont
  {Affleck}, \citenamefont {Oshikawa},\ and\ \citenamefont
  {Saleur}}]{potts_bcft98}%
  \BibitemOpen
  \bibfield  {author} {\bibinfo {author} {\bibfnamefont {Ian}\ \bibnamefont
  {Affleck}}, \bibinfo {author} {\bibfnamefont {Masaki}\ \bibnamefont
  {Oshikawa}}, \ and\ \bibinfo {author} {\bibfnamefont {Hubert}\ \bibnamefont
  {Saleur}},\ }\bibfield  {title} {\enquote {\bibinfo {title} {Boundary
  critical phenomena in the three-state potts model},}\ }\href {\doibase
  10.1088/0305-4470/31/28/003} {\bibfield  {journal} {\bibinfo  {journal}
  {Journal of Physics A: Mathematical and General}\ }\textbf {\bibinfo {volume}
  {31}},\ \bibinfo {pages} {5827} (\bibinfo {year} {1998})}\BibitemShut
  {NoStop}%
\bibitem [{\citenamefont {Vanhove}\ \emph {et~al.}(2022)\citenamefont
  {Vanhove}, \citenamefont {Lootens}, \citenamefont {Tu},\ and\ \citenamefont
  {Verstraete}}]{potts_dcft22}%
  \BibitemOpen
  \bibfield  {author} {\bibinfo {author} {\bibfnamefont {Robijn}\ \bibnamefont
  {Vanhove}}, \bibinfo {author} {\bibfnamefont {Laurens}\ \bibnamefont
  {Lootens}}, \bibinfo {author} {\bibfnamefont {Hong-Hao}\ \bibnamefont {Tu}},
  \ and\ \bibinfo {author} {\bibfnamefont {Frank}\ \bibnamefont {Verstraete}},\
  }\bibfield  {title} {\enquote {\bibinfo {title} {Topological aspects of the
  critical three-state potts model},}\ }\href {\doibase
  10.1088/1751-8121/ac68b1} {\bibfield  {journal} {\bibinfo  {journal} {Journal
  of Physics A: Mathematical and Theoretical}\ }\textbf {\bibinfo {volume}
  {55}},\ \bibinfo {pages} {235002} (\bibinfo {year} {2022})}\BibitemShut
  {NoStop}%
\bibitem [{\citenamefont {Fendley}\ \emph {et~al.}(2009)\citenamefont
  {Fendley}, \citenamefont {Fisher},\ and\ \citenamefont
  {Nayak}}]{Fendley_2009}%
  \BibitemOpen
  \bibfield  {author} {\bibinfo {author} {\bibfnamefont {Paul}\ \bibnamefont
  {Fendley}}, \bibinfo {author} {\bibfnamefont {Matthew~P.A.}\ \bibnamefont
  {Fisher}}, \ and\ \bibinfo {author} {\bibfnamefont {Chetan}\ \bibnamefont
  {Nayak}},\ }\bibfield  {title} {\enquote {\bibinfo {title} {Boundary
  conformal field theory and tunneling of edge quasiparticles in non-abelian
  topological states},}\ }\href {\doibase 10.1016/j.aop.2009.03.005} {\bibfield
   {journal} {\bibinfo  {journal} {Annals of Physics}\ }\textbf {\bibinfo
  {volume} {324}},\ \bibinfo {pages} {1547--1572} (\bibinfo {year}
  {2009})}\BibitemShut {NoStop}%
\bibitem [{\citenamefont {Cho}\ \emph {et~al.}(2017{\natexlab{a}})\citenamefont
  {Cho}, \citenamefont {Ludwig},\ and\ \citenamefont {Ryu}}]{Cho17b}%
  \BibitemOpen
  \bibfield  {author} {\bibinfo {author} {\bibfnamefont {Gil~Young}\
  \bibnamefont {Cho}}, \bibinfo {author} {\bibfnamefont {Andreas W.~W.}\
  \bibnamefont {Ludwig}}, \ and\ \bibinfo {author} {\bibfnamefont {Shinsei}\
  \bibnamefont {Ryu}},\ }\bibfield  {title} {\enquote {\bibinfo {title}
  {Universal entanglement spectra of gapped one-dimensional field theories},}\
  }\href {\doibase 10.1103/physrevb.95.115122} {\bibfield  {journal} {\bibinfo
  {journal} {Physical Review B}\ }\textbf {\bibinfo {volume} {95}} (\bibinfo
  {year} {2017}{\natexlab{a}}),\ 10.1103/physrevb.95.115122}\BibitemShut
  {NoStop}%
\bibitem [{\citenamefont {Cho}\ \emph {et~al.}(2017{\natexlab{b}})\citenamefont
  {Cho}, \citenamefont {Shiozaki}, \citenamefont {Ryu},\ and\ \citenamefont
  {Ludwig}}]{Cho17c}%
  \BibitemOpen
  \bibfield  {author} {\bibinfo {author} {\bibfnamefont {Gil~Young}\
  \bibnamefont {Cho}}, \bibinfo {author} {\bibfnamefont {Ken}\ \bibnamefont
  {Shiozaki}}, \bibinfo {author} {\bibfnamefont {Shinsei}\ \bibnamefont {Ryu}},
  \ and\ \bibinfo {author} {\bibfnamefont {Andreas W~W}\ \bibnamefont
  {Ludwig}},\ }\bibfield  {title} {\enquote {\bibinfo {title} {Relationship
  between symmetry protected topological phases and boundary conformal field
  theories via the entanglement spectrum},}\ }\href {\doibase
  10.1088/1751-8121/aa7782} {\bibfield  {journal} {\bibinfo  {journal} {Journal
  of Physics A: Mathematical and Theoretical}\ }\textbf {\bibinfo {volume}
  {50}},\ \bibinfo {pages} {304002} (\bibinfo {year}
  {2017}{\natexlab{b}})}\BibitemShut {NoStop}%
\bibitem [{\citenamefont {Cardy}(2017)}]{Cardy17}%
  \BibitemOpen
  \bibfield  {author} {\bibinfo {author} {\bibfnamefont {John}\ \bibnamefont
  {Cardy}},\ }\bibfield  {title} {\enquote {\bibinfo {title} {Bulk
  renormalization group flows and boundary states in conformal field
  theories},}\ }\href {\doibase 10.21468/scipostphys.3.2.011} {\bibfield
  {journal} {\bibinfo  {journal} {{SciPost} Physics}\ }\textbf {\bibinfo
  {volume} {3}} (\bibinfo {year} {2017}),\
  10.21468/scipostphys.3.2.011}\BibitemShut {NoStop}%
\bibitem [{\citenamefont {Fromholz}\ \emph {et~al.}(2019)\citenamefont
  {Fromholz}, \citenamefont {Capponi}, \citenamefont {Lecheminant},
  \citenamefont {Papoular},\ and\ \citenamefont {Totsuka}}]{Fromholz_2019}%
  \BibitemOpen
  \bibfield  {author} {\bibinfo {author} {\bibfnamefont {P.}~\bibnamefont
  {Fromholz}}, \bibinfo {author} {\bibfnamefont {S.}~\bibnamefont {Capponi}},
  \bibinfo {author} {\bibfnamefont {P.}~\bibnamefont {Lecheminant}}, \bibinfo
  {author} {\bibfnamefont {D.~J.}\ \bibnamefont {Papoular}}, \ and\ \bibinfo
  {author} {\bibfnamefont {K.}~\bibnamefont {Totsuka}},\ }\bibfield  {title}
  {\enquote {\bibinfo {title} {Haldane phases with ultracold fermionic atoms in
  double-well optical lattices},}\ }\href {\doibase 10.1103/physrevb.99.054414}
  {\bibfield  {journal} {\bibinfo  {journal} {Physical Review B}\ }\textbf
  {\bibinfo {volume} {99}} (\bibinfo {year} {2019}),\
  10.1103/physrevb.99.054414}\BibitemShut {NoStop}%
\bibitem [{\citenamefont {Capponi}\ \emph {et~al.}(2020)\citenamefont
  {Capponi}, \citenamefont {Fromholz}, \citenamefont {Lecheminant},\ and\
  \citenamefont {Totsuka}}]{Capponi_2020}%
  \BibitemOpen
  \bibfield  {author} {\bibinfo {author} {\bibfnamefont {S.}~\bibnamefont
  {Capponi}}, \bibinfo {author} {\bibfnamefont {P.}~\bibnamefont {Fromholz}},
  \bibinfo {author} {\bibfnamefont {P.}~\bibnamefont {Lecheminant}}, \ and\
  \bibinfo {author} {\bibfnamefont {K.}~\bibnamefont {Totsuka}},\ }\bibfield
  {title} {\enquote {\bibinfo {title} {Symmetry-protected topological phases in
  a two-leg $\text{SU}(n)$ spin ladder with unequal spins},}\ }\href {\doibase
  10.1103/PhysRevB.101.195121} {\bibfield  {journal} {\bibinfo  {journal}
  {Phys. Rev. B}\ }\textbf {\bibinfo {volume} {101}},\ \bibinfo {pages}
  {195121} (\bibinfo {year} {2020})}\BibitemShut {NoStop}%
\bibitem [{fut()}]{future_qt}%
  \BibitemOpen
  \href@noop {} {}\bibinfo {note} {Prembabu et al., to appear}\BibitemShut
  {NoStop}%
\bibitem [{\citenamefont {Verresen}\ \emph {et~al.}(2017)\citenamefont
  {Verresen}, \citenamefont {Moessner},\ and\ \citenamefont
  {Pollmann}}]{Verresen17}%
  \BibitemOpen
  \bibfield  {author} {\bibinfo {author} {\bibfnamefont {Ruben}\ \bibnamefont
  {Verresen}}, \bibinfo {author} {\bibfnamefont {Roderich}\ \bibnamefont
  {Moessner}}, \ and\ \bibinfo {author} {\bibfnamefont {Frank}\ \bibnamefont
  {Pollmann}},\ }\bibfield  {title} {\enquote {\bibinfo {title}
  {One-dimensional symmetry protected topological phases and their
  transitions},}\ }\href {\doibase 10.1103/PhysRevB.96.165124} {\bibfield
  {journal} {\bibinfo  {journal} {Phys. Rev. B}\ }\textbf {\bibinfo {volume}
  {96}},\ \bibinfo {pages} {165124} (\bibinfo {year} {2017})}\BibitemShut
  {NoStop}%
\bibitem [{\citenamefont {Kj\"all}\ \emph {et~al.}(2013)\citenamefont
  {Kj\"all}, \citenamefont {Zaletel}, \citenamefont {Mong}, \citenamefont
  {Bardarson},\ and\ \citenamefont {Pollmann}}]{Kjaell13}%
  \BibitemOpen
  \bibfield  {author} {\bibinfo {author} {\bibfnamefont {Jonas~A.}\
  \bibnamefont {Kj\"all}}, \bibinfo {author} {\bibfnamefont {Michael~P.}\
  \bibnamefont {Zaletel}}, \bibinfo {author} {\bibfnamefont {Roger S.~K.}\
  \bibnamefont {Mong}}, \bibinfo {author} {\bibfnamefont {Jens~H.}\
  \bibnamefont {Bardarson}}, \ and\ \bibinfo {author} {\bibfnamefont {Frank}\
  \bibnamefont {Pollmann}},\ }\bibfield  {title} {\enquote {\bibinfo {title}
  {Phase diagram of the anisotropic spin-2 xxz model: Infinite-system density
  matrix renormalization group study},}\ }\href {\doibase
  10.1103/PhysRevB.87.235106} {\bibfield  {journal} {\bibinfo  {journal} {Phys.
  Rev. B}\ }\textbf {\bibinfo {volume} {87}},\ \bibinfo {pages} {235106}
  (\bibinfo {year} {2013})}\BibitemShut {NoStop}%
\end{thebibliography}%

\onecolumngrid
\appendix
\setcounter{figure}{0}
\renewcommand\thefigure{\thesection.\arabic{figure}}

\newpage
\begin{center}
{\large \textbf{Supplemental Material}}
\end{center}

\title{Supplemental Material: Boundary deconfined quantum criticality at transitions\\between symmetry-protected topological chains}

\author{Saranesh Prembabu}
\affiliation{Department of Physics, Harvard University, Cambridge, Massachusetts 02138, USA}

\author{Ryan Thorngren}
\affiliation{Kavli Institute of Theoretical Physics, University of California, Santa Barbara, California 93106, USA}

\author{Ruben Verresen}
\affiliation{Department of Physics, Harvard University, Cambridge, MA 02138, USA}

\maketitle

\onecolumngrid
\appendix
\setcounter{figure}{0}


\section{Mapping to single Potts chain \label{app:mapping}}

The nonlocal unitary map of Fig. 3 of the main text from the cluster to Potts model is made explicit in Table \ref{tab:odd_odd_mapping}. 

\begin{table}[h!]
    \centering

    \begin{minipage}[b]{0.3\linewidth}\centering
    \begin{tabular} {|c|c|c|}
    \hline
     Potts${}$ & Cluster \\ \hline \hline
     $\tilde{X}_{j} \tilde{X}_{1-j}^{ }$ & $X_{2 \le 2j \le 2N}$ \\ \hline
     $\tilde{Z}_{j}^{\dagger} \tilde{Z}_{1-j}^{\dagger} $& $Z_{2 \le 2j \le 2N}$ \\ \hline
      $\tilde{Z}_{j-1}^{} \tilde{Z}_{2-j}^{\dagger} \tilde{Z}_{j}^{ \dagger} \tilde{Z}_{1-j}^{}$ & $X_{3 \le 2j-1 \le 2N-1 }$ \\ \hline
      $W\prod_{k=j}^{N} \tilde{X}_{k}^{\dagger} \tilde{X}_{1-k}^{ }$ & $Z_{1 \le 2j-1 \le 2N+1 }$ \\ \hline
      $\tilde{Z}_{1}^{\dagger} \tilde{Z}_0^{ }$ & $X_1$ \\ \hline
      $Y^\dagger \tilde{Z}^{}_{N}\tilde{Z}^{\dagger}_{N+1} $ & $X_{2N+1}$ \\
    \hline 
    \end{tabular}
    \end{minipage}
    \hspace{0.5cm}
    \begin{minipage}[b]{0.3\linewidth}
    \centering
    
    \begin{tabular}{|c|c|}
    \hline
     Potts${}$ & Cluster \\ \hline \hline
      $\tilde{X}_{0 \ge 1-j \ge 1-N}^{B}$& $Z_{2j-1}^\dagger X_{2j}^\dagger Z_{2j+1}$ \\ \hline
     $ \tilde{X}_{1 \le j \le N}^{}$& $Z_{2j-1} X_{2j}^\dagger Z_{2j+1}^\dagger$ \\ \hline
      $\tilde{Z}_{1 \le j \le N}^{}$&$X_1 X_3 \ldots X_{2j-1} Z_{2j}$ \\ \hline
      $\tilde{Z}_{0 \ge 1-j \ge 1-N}^{}$& $ X_1^\dagger X_3^\dagger \ldots X_{2j-1}^\dagger Z_{2j}$\\ \hline
     $Y$ & $\prod_{j=0}^{N} X_{2j+1}^\dagger $ \\ \hline
      $W$ & $Z_{2N+1}$ \\ \hline
      \end{tabular}
    \end{minipage}
    \hspace{0.5cm}
    \begin{minipage}[b]{0.3\linewidth}
    \centering
    \begin{tabular}{|c|c|}
      \hline
      Potts${}$ & Cluster \\ \hline \hline
      $\prod_{j=1}^{2N} \tilde{X}^{}_{j} $ & $\prod_{j=1}^{N} X_{2j}$ \\ \hline

      $ Y^\dagger$ &$ \prod_{j=0}^{N} X_{2j+1}$ \\ \hline
      $\prod_{j=1}^{N} \tilde{X}^{}_{j} \tilde{X}^{\dagger}_{1-j}$ & $Z_1 Z^\dagger_{2N+1}$ \\ \hline
    
    \end{tabular}
    \end{minipage}
    
    \caption{\textbf{Mapping from cluster to Potts:} An exact equivalence between the cluster model on an open chain with odd-site endpoints and the Potts model. Here the cluster model is defined on all integer sites $1\le j \le 2N+1$. The Potts chain is defined on integer sites $1 \le j \le 2N$ with periodic boundary conditions $j \equiv j+2N$; the Potts chain couples to an additional qutrit described by clock and shift matrices $Y$ and $W$ representing a gauged degree of freedom.
    \label{tab:odd_odd_mapping}}
\end{table}

\section{Boundary CFT and Defect CFT Analysis  \label{app:bcft}}

\begin{figure}
    \centering
    \begin{tikzpicture}
    \node at (-8,0) {\includegraphics[width =8cm]{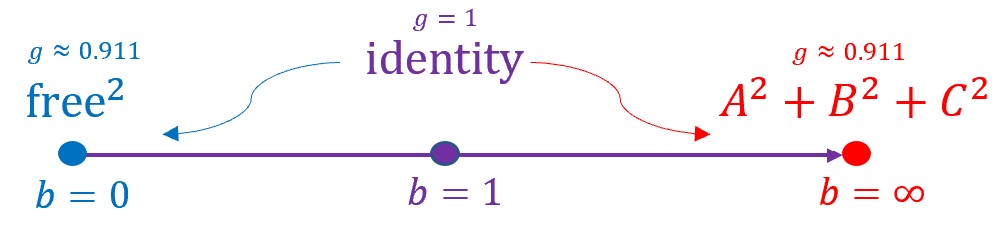}};
    \node at (0,0) { \includegraphics[width = 4 cm]{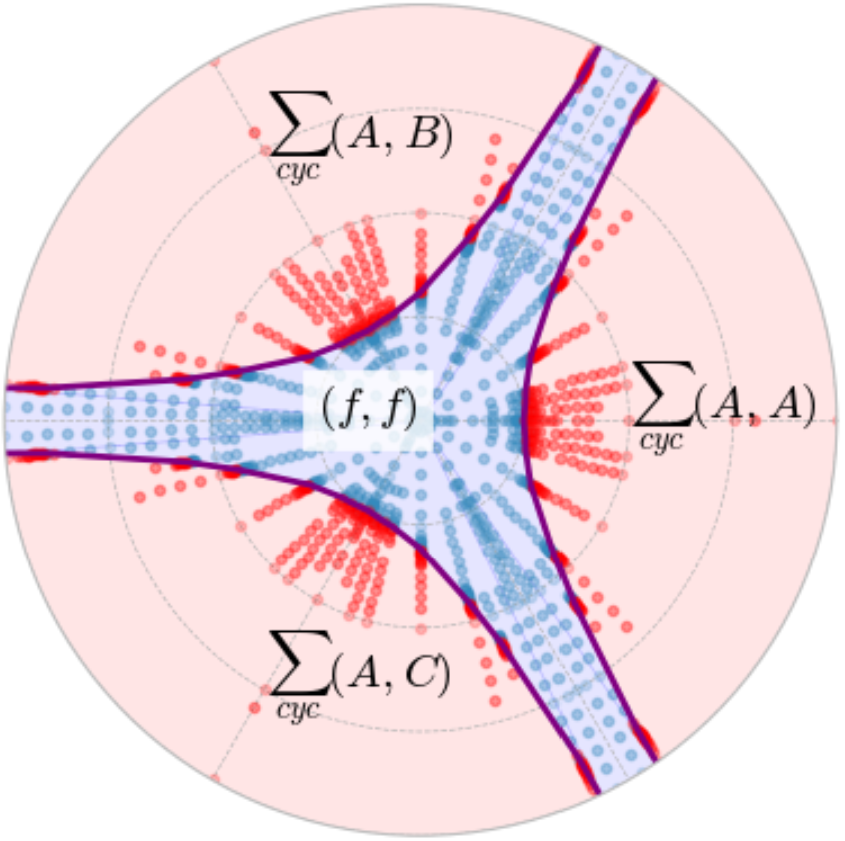}};
    \node at (0,2.2) {$\textrm{Im}(b)$};
    \node at (2.5,0) {$\textrm{Re}(b)$};
    \node at (-12,2.2) {(a)};
    \node at (-2.5,2.2) {(b)};
    \end{tikzpicture}
    \caption{\textbf{The conformal defect that arises by changing the strength $b$ of one bond in the Potts model}. (a) For $b\ge 0$ there are three different defects in the field theory. At $b=1$ there is no defect (identity defect, shown in purple), which is unstable to local perturbation. Tuning $b$  below or above 1 triggers an RG flow to the stable defects $(\textrm{free}, \textrm{free})$ (shown in blue)  or  $\sum_{cyc}(A,A) \equiv (A,A) +(B,B)+(C,C)$ (shown in red) respectively. (b) Conformal defects for complex $b = |b|e^{i\theta}$ computed numerically (colored dots) from long range order of cluster order parameters. A topological twist defect relates $\theta$ and $\theta+2\pi/3$.
    }
    \label{fig:b_phase_diagram}
    \label{fig:imaginary_b}
\end{figure}
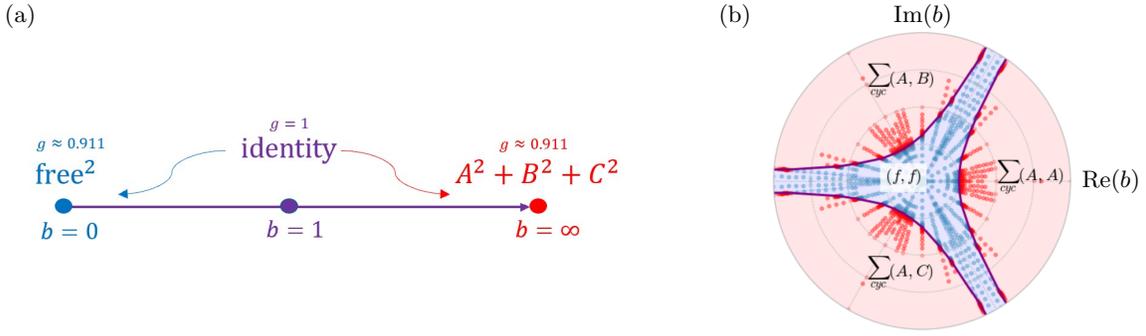

Thanks to the above mapping, instead of directly identifying the boundary conditions of the gapless cluster model, we can indirectly characterize them using some well-known defect descriptions of the Potts model.  
The low energy description of the bond-strength defect $b$ in the Potts Hamiltonian takes three universal forms for different regimes of $b$, shown in Fig.\ref{fig:b_phase_diagram} and Table \ref{tab:b_defects}. For each regime, we  summarize the important CFT results.

\[ \cdots - \tilde Z_{-2}\tilde Z_{-1}^\dagger - \tilde X_{-1} -\tilde Z_{-1}\tilde Z_0^\dagger - \tilde X_0 -\color{red}b\tilde Z_0\tilde Z_1^\dagger \color{black} - \tilde X_1 - \tilde Z_1\tilde Z_2^\dagger- \tilde X_2 - \tilde Z_2\tilde Z_3^\dagger - \tilde X_3 \cdots + \textrm{h.c.} \]

\begin{table}[h!]
    \centering
    \begin{tabular}{|c||c|c|c|} \hline
       $b$  &  $[0,1)$ & $1 $ & $(1, \infty)$ \\\hline \hline
         RG Flow & $b\to 0$ & $b=1$ & $b\to\infty$ \\ \hline
         Potts Defect & $({\rm free}, {\rm free})$ & No defect & $(A,A)+(B,B)+(C,C)$\\\hline
         Cluster Boundary & $o$-SSB & DQCP & $e$-SSB \\ \hline
    \end{tabular}
    \caption{Different regimes of boundary perturbation strength $b$}
    \label{tab:b_defects}
\end{table}

\subsection{Boundary-SSB}

The $o$-SSB and $e$-SSB conformal boundary conditions map to each other by the Potts model's Krammers Wannier duality. 
Without loss of generality let us consider $o$-SSB. The underlying theory has three different conformal vacua which each return nonzero vev for $\bZ_3^o$-charged lattice order parameters. We can denote the local $\mathcal{W}_3^A \otimes \mathcal{W}_3^B$-primary boundary CFT operators as $\psi_{A}$ and $\psi_{B}$, with $\bZ_3^e$ charge $\omega$ and conformal dimension $2/3$, as well as all fusion products of these. 
The most relevant neutral boundary operator $\psi_A \psi_B^\dagger$ has dimension $4/3>1$, explaining the perturbative stability of this boundary condition and the finite size gap scaling exponent $2(4/3)-1 = 5/3$. Meanwhile, the relevant boundary operators $\psi_{A/B}$ are charged and play the role of boundary disorder operators.

\subsection{DQCP}

The theory with the DQCP boundary on both sides has a unique conformal vacuum and has local boundary CFT operators in one-to-one correspondence with the Potts model's local bulk operators and nonlocal $\bZ_3$-twisted-sector operators. The most relevant neutral boundary operator $\epsilon$ has dimension $\Delta_{\epsilon} = 4/5$, leading to the perturbative instability of the DQCP boundary condition upon changing $b$ above or below $1$. Furthermore, the $e$-SSB's and $o$-SSB's charged order parameters correspond to the DQCP's charged operators $\sigma$ and $\mu$ respectively and are manifestly interchanged by the Potts Krammers-Wannier order-disorder duality. In agreement with the numerics of Fig. 1 (d) of the main text, as $b\to 1$ boundary order parameters vanish as  $\propto (b-1)^{\frac{\Delta_{\sigma}}{1-\Delta_{\epsilon}}} = (b-1)^{2/3}$ from the $e$-SSB regime and $\propto (1-b)^{\frac{\Delta_{\mu}}{1-\Delta_{\epsilon}}} = (1-b)^{2/3}$ from the $o$-SSB regime. From the \textit{gapped} SPT phases as $s\to 0$, both boundary order parameters vanish as $\propto |s|^{\frac{\Delta_{\sigma,\mu}}{2-\Delta_s}} = |s|^{1/9}$ where $\Delta_s=4/5$ is the dimension of the bulk operator perturbing to the gapped SPT.

\section{Boundary State Analysis \label{app:cluster_boundary_state}}

Boundary conditions of the cluster model can be expressed in terms of boundary states living in the cluster model's bulk Hilbert space \cite{cardyBCFT}. Here we write down explicitly some of these boundary states. For convenience states are written using the notation of the Potts$^2$ model, keeping in mind the actual Hilbert space is its antidiagonal $\bZ_3$ orbifold.

\begin{equation}
\begin{split}
    |\textrm{o-ssb}\rangle 
    & = 3\sqrt{3}N^2 \left( |0\>\>-|3\>\> - \sqrt{\varphi}|2/5\>\> + \sqrt{\varphi} |7/5\>\>\right)_A \left( |0\>\>-|3\>\> - \sqrt{\varphi}|2/5\>\> + \sqrt{\varphi} |7/5\>\>\right)_B\\
    |\textrm{e-ssb}\rangle 
    &=  3\sqrt{3}N^2 \left( |0\>\>+|3\>\> + \sqrt{\varphi}|2/5\>\> + \sqrt{\varphi} |7/5\>\>\right)_A \left( |0\>\>+|3\>\> + \sqrt{\varphi}|2/5\>\> + \sqrt{\varphi} |7/5\>\>\right)_B\\
    |\textrm{DQCP}\rangle
    &= \sqrt{3}\sum_{h} \sum_{J} |h, J\rangle_A |h, J \rangle_B \\
\end{split}
\end{equation}

Here $\phi$ is the golden ratio and $N=\left( \frac{5-\sqrt{5}}{30} \right)^{1/4}$. The notation $|h,J\rangle_{A/B}$ refers schematically to the $J$-th descendant of Potts bulk primary $h$, while $|h \>\>_{A/B}$ refers to the Potts model Ishibashi state.  $|\textrm{DQCP}\rangle$ is a simple boundary condition, while $|\textrm{o/e-SSB}\rangle$ each decompose into a sum of three simple $\bZ_3^{o/e}$-breaking boundary states. For example $|\textrm{e-ssb}\>$ is the sum of three distinct boundary states $|\textrm{e-}a\>$, where

\begin{equation}
    \begin{split}
    |\textrm{e-}a\rangle &\equiv \frac{1}{\sqrt{3}} \sum_{b=0}^2 |\omega^{a+b}\rangle_A |\omega^{a-b}\rangle_B \\
    \qquad
    |\omega^a\rangle &\equiv N \left( |0\>\>+|3\>\> + \omega^a |\psi \>\> + \omega^{-a} |\psi^\dagger \> \> + \sqrt{\varphi} \left(|2/5 \>\>+|7/5 \>\> + \omega^a |\sigma \>\> + \omega^{-a} |\sigma^\dagger\>\> \right) \right) \\
    \end{split}
\end{equation}

\section{Adjacent Gapped Phases \label{app:nearbyphases}}

Our gapless model $H_{\omega}+H_{\bar{\omega}}$ is not adjacent to a trivial gapped phase. Its nearby phases are the two gapped SPT phases, the two gapped SSB phases breaking the odd or even $\bZ_3$ subgroups, and transitions between these. These can be identified by considering relevant bulk CFT operators \cite{RJM21, Roberts_conversation, Whitsitt_2018}.
Imposing translation and charge conjugation symmetries (in addition to $\bzz$) stabilizes a one-parameter family of models containing $H_{\omega}+H_{\bar{\omega}}$, a first order transition between the gapped SSB phases, and a first order transition between the gapped SPT phases.
Adding the bulk odd (even) SSB perturbation to a semi-infinite region of the infinite gapless $H_{\omega}+H_{\bar{\omega}}$ chain results in the $o$-SSB ($e$-SSB) boundary condition.

\textit{}

\end{document}